\documentclass[11pt,a4paper,oneside,dvipsnames]{article}
\usepackage[table]{xcolor}
\definecolor{AQI_1}{HTML}{b3fca6}
\definecolor{AQI_2}{HTML}{7bfb4c}
\definecolor{AQI_3}{HTML}{67cc3c}
\definecolor{AQI_4}{HTML}{ffff54}
\definecolor{AQI_5}{HTML}{f7d147}
\definecolor{AQI_6}{HTML}{f19e39}
\definecolor{AQI_7}{HTML}{ed6d69}
\definecolor{AQI_8}{HTML}{ea3423}
\definecolor{AQI_9}{HTML}{8c1911}
\definecolor{AQI_10}{HTML}{bf3ff6}

\usepackage{graphicx}
\usepackage{rotating}
\usepackage{wrapfig}
\usepackage{placeins}

\usepackage{tabularx}
\usepackage{booktabs} % for better looking tables

\newcolumntype{+}{>{\global\let\currentrowstyle\relax}}
\newcolumntype{^}{>{\currentrowstyle}}

\usepackage{sidecap}

\usepackage[top=1in]{geometry}

\usepackage{latexsym}
\usepackage{graphicx}
\usepackage{multicol,multirow}
\usepackage{amsmath,amssymb,amsfonts}
\usepackage{mathrsfs}
\usepackage{amsthm}
\usepackage{rotating}
\usepackage{appendix}
\usepackage{ifpdf}
\usepackage[T1]{fontenc}
\usepackage{times}
\usepackage{sourcesanspro}
\usepackage{newtxmath}
\usepackage{textcomp}%
\usepackage{xcolor}%
\usepackage{csvsimple}
\usepackage{adjustbox}
\usepackage{xr}
\usepackage{soul}

\usepackage{xr-hyper}
\usepackage{hyperref}
\usepackage{siunitx}
\usepackage{authblk}

\usepackage{subcaption}
\usepackage[utf8]{inputenc}

\usepackage{array}

\usepackage{listings}

\definecolor{darkgreen}{rgb}{0.0, 0.5, 0.0}

\lstdefinestyle{mypy}{
    language=Python,
    basicstyle=\ttfamily\small,
    commentstyle=\color{olive},
    keywordstyle=\color{blue},
    numberstyle=\tiny\color{gray},
    numbers=left,
    stringstyle=\color{red},
    showstringspaces=false,
}

\hypersetup{
    colorlinks = true,
    linkcolor = blue,
    filecolor = magenta,      
    urlcolor = cyan,
}

\DeclareUnicodeCharacter{2080}{\textsubscript{0}}
\DeclareUnicodeCharacter{2081}{\textsubscript{1}}
\DeclareUnicodeCharacter{2082}{\textsubscript{2}}
\DeclareUnicodeCharacter{2083}{\textsubscript{3}}
\DeclareUnicodeCharacter{2084}{\textsubscript{4}}
\DeclareUnicodeCharacter{2085}{\textsubscript{5}}
\DeclareUnicodeCharacter{2093}{\textsubscript{x}}

\usepackage{geometry}

\newgeometry{vmargin={15mm}, hmargin={12mm,17mm}}   % set the margins

\usepackage{booktabs}
\usepackage{pgfplotstable} % Generates table from .csv

\definecolor{reddish}{HTML}{FBB4AE}
\definecolor{blueish}{HTML}{B3CDE3}
\definecolor{magentish}{HTML}{FF00AA}
\definecolor{greenish}{HTML}{a1d99b}

\makeatletter
\newcommand*{\addFileDependency}[1]{% argument=file name and extension
  \typeout{(#1)}
  \@addtofilelist{#1}
  \IfFileExists{#1}{}{\typeout{No file #1.}}
}
\makeatother

\title{Environmental Insights: Democratizing Access to Ambient Air Pollution Data and Predictive Analytics with an Open-Source Python Package}
\author{
  Liam J. Berrisford$^{1,2,3,*}$\\
  \texttt{l.berrisford@exeter.ac.uk}
 \and
  Ronaldo Menezes$^{1,4}$\\
  \texttt{r.menezes@exeter.ac.uk}
}

\date{%
    $^1$ BioComplex Laboratory, Department of Computer Science, University of Exeter, England\\%
    $^2$ Department of Mathematics, University of Exeter, England\\
    $^3$ UKRI Centre for Doctoral Training in Environmental Intelligence, University of Exeter, England\\%
    $^4$ Department of Computer Science, Federal University of Ceará, Fortaleza, Brazil\\%
    $^{*}$ Corresponding Author\\[2ex]%
    \today
}

\begin{document}

\maketitle

\begin{abstract}
    Ambient air pollution is a pervasive issue with wide-ranging effects on human health, ecosystem vitality, and economic structures. Utilizing data on ambient air pollution concentrations, researchers can perform comprehensive analyses to uncover the multifaceted impacts of air pollution across society. To this end, we introduce {\em Environmental Insights}, an open-source Python package designed to democratize access to air pollution concentration data. This tool enables users to easily retrieve historical air pollution data and employ a Machine Learning model for forecasting potential future conditions. Moreover, {\em Environmental Insights} includes a suite of tools aimed at facilitating the dissemination of analytical findings and enhancing user engagement through dynamic visualizations. This comprehensive approach ensures that the package caters to the diverse needs of individuals looking to explore and understand air pollution trends and their implications. 
    \\
    {\bfseries Code Repository Clickable Link}: \href{https://github.com/berrli/Environmental-Insights?tab=readme-ov-file}{Environmental Insights Github Home Page}\\ 
    {\bfseries Data and Model Repository Clickable Link}: \href{https://drive.google.com/drive/folders/18ZLO8XqtFp3c4WrUJVfSH0fmAXFmL8il?usp=sharing}{Environmental Insights Google Drive Folder}\\
\end{abstract}

\clearpage
\section{Introduction}
Air pollution impacts all societal sectors, generating widespread interest in air pollution concentration data among diverse stakeholders, including academic and research institutions, government agencies, environmental NGOs, the private sector, the medical community, and the general public. Extensive research has been conducted on predicting air pollution concentrations using various modelling frameworks \cite{Henze:2007:GeosChem, hoek:2008:review, freeman:2018:ForecastingAirPollutionConcentration, tao:2019:ForecastingAirPollutionConcentration2, harishkumar:2020:ForecastingAirPollutionConcentration3, van:2019:MissingLocationAirPollutionEstimationSmallArea, chen:2021:MissingLocationAirPollutionMonthly, he:2023:AirPollutionLeaveOneOutValidationDaily, li:2020:AirPollutionLeaveOneOutValidationDaily2}. However, leveraging air pollution concentration data should not be seen as a unilateral process where predictions are simply delivered to stakeholders without further engagement. Instead, an iterative approach that considers the practical use and outcomes of these predictions is crucial for refining and directing future research concerning air pollution.

In response to this need, our work introduces \textit{Environmental Insights}, an open-source Python package designed to facilitate active engagement with air pollution issues. This package enables stakeholders to download, analyse, and visualise air pollution concentration data, thereby offering a unified platform for exploring potential air pollution futures. \textit{Environmental Insights} aims to disseminate and democratise access to air pollution data, breaking down barriers for individuals and communities without extensive resources or technical expertise. By empowering a broader audience to engage with air pollution data, the package also seeks to amplify public pressure on policymakers for meaningful air quality improvements in areas of significant concern to the community.

\section{Use Cases For Air Pollution Concentration Data}

Air pollution concentration data serves as a vital foundation across a wide spectrum of applications, addressing critical needs in health assessments, environmental protection, economic analysis, and fundamental research. This data is instrumental not only for immediate health and environmental concerns but also for strategic urban planning, policy development, and enhancing community engagement. However, the importance of air pollution data transcends various domains, with stakeholders ranging from government entities involved in setting standards, regulations, and laws \cite{EUCOuncil:2008@AirPollutionConcentrationLimitDirective, Harrabin:2021:UKSuedByEUAirPollutionLimits, UKGov:2007:Euro6EmissionsStandard}, to urban planners \cite{anderson:1971:AirPollutionHousingValue, liu:2015:AirPollutionCommuting, fischer:2018:AIrPollutionGreenspaceAssessment} and parties in legal proceedings \cite{Laville:2020:EllaAirPollutionDeathLondon, thakrar:2020:AttributableAirPollutionConcentrations} who leverage this data for a multitude of purposes. The accuracy and high temporal and spatial resolution of this data are of clear value to various stakeholders, highlighting the necessity for tools that facilitate its analysis and visualisation. Such tools are crucial for promoting the agendas of each stakeholder, a theme further explored in the subsequent sections of this paper.

\subsection*{Human Health Assessments}

Data on the levels and types of air pollutants in the atmosphere allows a range of stakeholders, including researchers, policymakers, and public health officials, to engage in human health assessments, enabling informed decisions concerning public health interventions, regulations, and public awareness campaigns.

The first step is to make an exposure assessment, understanding the concentration, duration, and frequency of air pollutant exposure, to provide an estimate for an individual's inhalation of said pollutants \cite{USEPA:1992:GuidelinesForExposureAssessment}. With an assessment of the exposure, the relation to the health impacts is then conducted.

However, while the individual effects of air pollution are critical, a key focus of national agencies and research is on understanding more general impacts on the population and subpopulations to enable the subsequent design of interventions to protect vulnerable populations \cite{zou:2009:AirPollutionExposureMethodsEpidemiological}, with extensive studies in areas from China \cite{chen:2017:ChinaEpidemiologicalStudy} to the UK \cite{atkinson:2016:UKEpidemiologicalStudy}. Epidemiological studies are particularly critical when broad changes in air pollution are being observed, such as during COVID-19 \cite{konstantinoudis:2021:UKEpidemiologicalStudyCovid19}.

The form of interventions designed through these assessments is twofold. Understanding the implications of air pollution with real-time concentration data allows for emergency mitigations to be implemented during high air pollution episodes \cite{duan:2023:EmergencyMitigationForAirPollutionExtreme}. Secondly, long-term health service needs can be understood from future projections of ailments from air pollution exposure assessments \cite{lipfert:1993:LongTermHealthServiceNeedAssessments}.

\subsection*{Ecosystem Health Assessment}

Air pollution can also affect the health of ecosystems and the environment more generally. Human health and environmental health are intertwined, with both impacting one another \cite{lovell:2018:HumanHealthAndEnvironment}. Understanding the stress on the environment that air pollution can cause is vital. Air pollutants such as O$_3$, SO$_2$, and NO$_x$ can cause damage to plants, such as leaf injury and stunted growth \cite{molnar:2020:AirPollutionVegtation}, with pressing concerns regarding reduced crop yields and food security \cite{tai:2014:AirPollutionOzoneFoodSecurity}. Further, air pollution can cause particular environmental phenomena, such as acid rain, from the emissions of SO$_2$ and NO$_x$ \cite{sivaramanan:2015:acidRainCauses}. Algal blooms and low oxygen levels in water bodies can be caused by NO$_x$, killing fish and other aquatic life, known as eutrophication \cite{glibert:2005:NO2eutrophication}. The pH and nutrient content of soil can experience degradation, reducing fertility and subsequent impacts on plant growth due to air pollution \cite{smith:2012:AirPollutionSoil}, with similar issues seen in water bodies \cite{maipa:2001:AirPollutionWaterQuality}. Particularly concerning is the situation of pollutants accumulating in organisms and being ingested by other organisms, with the concentrations increasing up the food chain. Eventually, these concentrations can become overwhelming and harm predators, including humans \cite{kampa:2008:FoodChainAirPollutionEscape}.

\subsection*{Fundamental Research}

So far, the discussion has focused primarily on downstream application areas of air pollution concentration data. However, it can also be used in fundamental research. The data is critical to understanding atmospheric chemical processes, including how air pollutants interact with each other \cite{zhu:2023:InterAirPollutionInteraction} and the atmosphere itself \cite{li:2017:AirPollutionBoundaryLayerInteraction}.

\subsection*{Economic Impact}

Air pollution can have serious economic consequences; concentration data is critical to evaluating and quantifying those impacts. Closely linked to human health assessment are estimates of healthcare costs due to air pollution. Air pollution concentrations can be correlated with hospital admissions \cite{schwartz:1996:AirPollutionHospitalAdmissions}, outpatient visits \cite{li:2017:AirPollutionOutpatientVists}, and sales of medications \cite{casas:2016:AirPollutionMedicationSales}, with long-term exposure related to extended medical treatments \cite{xu:2022:AirPollutionLongTermHealthCareCosts}. Furthermore, air pollution can impact labour productivity through absenteeism \cite{ponka:1990:AirPollutionAbsenteeism} and presenteeism \cite{goyal:2023:AirPollutionPresenteeism}. Air pollution can also affect revenue-generating activities such as tourism reduction during high air pollution episodes \cite{zhang:2020:AirPollutionTourism}. However, the reverse is also seen as true by some stakeholders, with air pollution an unavoidable consequence of economic growth \cite{Guardian:2023:BelgradeAirPollution}.

\subsection*{Public Engagement}

Air pollution concentration data is also of interest to non-professional stakeholders, where air pollution can impact their lives and communities. Some of these stakeholders include citizen communities, such as Mums for Lungs, raising awareness of the health effects of air pollution, particularly on children\footnote{\href{https://www.mumsforlungs.org/}{Mums for Lungs Webpage. (Last accessed: 05/03/2024)}}, and local communities such as Clean Air for Southall and Hayes\footnote{\href{https://www.breathelondon.org/community-groups/clean-air-for-southall-and-hayes}{Clean Air for Southall and Hayes (CASH). (Last accessed: 05/03/2024)}}. Some communities are focused on lobbying for policy changes, such as Friends of the Earth\footnote{\href{https://friendsoftheearth.uk/climate/air-pollution}{Friends of the Earth. (Last accessed: 05/03/2024)}}. Specialist groups, such as patients with ailments like Asthma+ Lung UK\footnote{\href{https://www.asthmaandlung.org.uk/about-us}{Asthma+Lung UK. (Last accessed:05/03/2024)}} are also keenly interested in air pollution concentration data.

\section{Existing Software}
\label{sec:existingSoftware}

Various software enables users to engage with air pollution concentration data and perform analysis. These can range from online portals, allowing for exploration and analysis, standalone pieces of software to engage with the data and software packages for various programming languages. 

The most accessible tools for analysing air pollution concentration data are online portals. For example, OpenAQ features a data explorer\footnote{\href{https://explore.openaq.org/}{OpenAQ Interactive Map. (Last accessed: 05/03/2024)}}. The map allows the user to click on existing air pollution monitoring stations and see basic statistics for the station, alongside downloading the data for the station in a standard format, such as comma-separated values (.csv). Other platforms also follow the same design, such as PurpleAir\footnote{\href{https://map.purpleair.com/}{PurpleAir Interactive Map. (Last accessed: 05/03/2024)}}. The data downloaded from these interactive maps can be used in other applications, such as geographic information system (GIS) software, to interrogate and perform more advanced analyses on the work, like QGIS \cite{QGIS:2023:GISSoftware}.

For more computer-savvy users, various software packages for programming languages are designed for air pollution. In R, the package {\em open-air} \cite{Carslaw:2012:OpenAir} provides access to real-world air pollution observations and tools to analyse air pollution, such as wind rose plots, air pollution roses, time variation, etc. {\em AirSensor} \cite{MazamaScience:2023:AirSensor} is a package that can give access to the low-cost PurpleAir sensors, with the tools in other packages, such as {\em open-air}, used for further analysis on top of the tools already provided within R. The R package {\em rdefra}~\cite{Ropensci:2023:RDefra} from the UK Department for Environment, Food and Rural Affairs allows access to a wealth of air pollution data maintained by the department. While some Python packages can access air pollution data, such as {\em py-openaq} \cite{Hagan:2023:PyOpenAQ}, and some packages can perform analysis on the data, such as {\em PyAir} \cite{Roubeyrie:2023:PyAir}, these are less well-established than the packages in R.

There is also more advanced open-source software capable of predicting air pollution concentrations. This software predominantly originates from national research institutions and predominantly employs atmospheric dispersion modelling approaches, such as {\em AEROMOD} \cite{cimorelli:2005:AEROMOD}, a dispersion model provided by the EPA, {\em HYSPLIT} by the NOAA Air Resources Laboratory (ARL) \cite{stein:2015:HYSPLIT}, {\em EPISODE} by the Norwegian Institute for Air Research (NILU) \cite{walker:1999:EPISODE}, and {\em WRF-Chem} \cite{peckham:2012:WRFChem} and {\em GEOS-Chem} \cite{Henze:2007:GeosChem}. The primary challenge with these models is their resource intensity and the extensive expertise required for effective operation, exemplified by {\em GEOS-Chem}, which demands 32GB of RAM as standard \footnote{\href{https://gchp.readthedocs.io/en/latest/getting-started/requirements.html}{{\em GEOS-Chem} Hardware Requirements. (Last accessed: 05/03/2024)}}.

As such, we observe a notable gap in the current offerings of software capable of modelling air pollution concentrations. None of the available software solutions offer the capability to predict air pollution concentrations from modified scenarios in a lightweight and user-friendly manner. A primary advantage of our preceding work is the computation speed afforded by a data-driven machine learning approach to estimating air pollution concentrations \cite{berrisford:2024:EstimatingAnnualAirPollution, berrisford:2024:EstimatingAirPollutionHourlyEngland, berrisford:2024:EstimatingAirPollutionHourlyGlobal}. Although the tools and software previously described are effective for examining air pollution concentration data, we regard them as static. In this work, we introduce a Python package, \textit{Environmental Insight}, leveraging the efficient performance characteristics of machine learning to provide an accessible model for estimating air pollution on consumer-grade laptops without the need for specialised knowledge. This approach renders air pollution estimation accessible to all stakeholders, enabling anyone to participate in air pollution concentration interventions. The remainder of this work details the various components of the package and illustrates its application in engaging stakeholders with air pollution-related issues, complemented by example use cases.

\section{Data}
\subsection{Accessing High Resolution Air Pollution Concentration Predictions}
\label{sec:dataAirPollutionConcentrations}

Accessing air pollution concentration data with the package is designed to be user-friendly. To access the data, users only need to specify latitude, longitude, and a timestamp. The system then returns the closest air pollution data point from the requested dataset. Accompanying this data, metadata regarding the proximity of the nearest data point is also provided, enabling users to make informed decisions about the relevance and applicability of the data for their specific needs.

Two underlying datasets achieve this. The first is the high-resolution hourly 1km$^2$ resolution data for the United Kingdom, and the second is the global dataset at hourly 0.25$^{\circ}$ resolution. Alongside accessing individual points, the user can access the complete dataset for a given timestamp, allowing for interpolation or further, more comprehensive data processing.

\noindent\textbf{Data Accessing Functions}

\begin{itemize}
    \item Function: \textit{air\_pollution\_concentration\_complete\_set\_real\_time\_united\_kingdom} Retrieve the complete predicted dataset for a given timestamp in the England dataset.
    \item Function: \textit{air\_pollution\_concentration\_nearest\_point\_real\_time\_united\_kingdom} Retrieve a single air pollution concentration data point predicted based on the England data, based on the closest point given by the latitude and longitude.
    \item Function \textit{air\_pollution\_concentration\_complete\_set\_real\_time\_global} Retrieve the complete predicted dataset for a given timestamp in the global dataset.
    \item Function \textit{air\_pollution\_concentration\_nearest\_point\_real\_time\_global} Retrieve a single air pollution concentration data point predicted based on the global data, based on the closest point given by the latitude and longitude.
\end{itemize}

\subsection{Distilling Data With Air Quality Index and Bands}

The main downstream output produced from air pollution concentration data is Air Quality Indexes (AQI), providing the public with clear information about air pollution levels. As one of the main outputs produced from air pollution concentration data, the package presented provides functions for creating an AQI.

There are a range of different air quality indexes across countries, such as the US EPA Air Quality Index (AQI) \cite{USEPA:1999:AQIFromPSI}, the Canadian Air Quality Health Index (AQHI) \cite{GovCanada:2021:AirQualityHealthIndexDefinition}, and the Chinese Air Quality Standards \cite{wang:2019:ChineseAirQualityStandards}, with some countries having indexes for particular subregions such as in Australia \cite{AustraliaGov:2023:AirQualityGuidelinesLinkTree}. More broadly, there is also the WHO Air Quality Guidelines \cite{WHO:2021:GlobalAQG}. While each has its merit, the index focused on for this package was the UK Daily Air Quality Index (DAQI) \cite{UKAIR:2023:DAQI}. The reason for focusing on the DAQI was that the highest resolution dataset was from the England, alongside the UK DAQI having a broad remit, not focusing solely on health outcomes, and supporting guidance for public dissemination, such as associated colours for each index. While it would be possible to include multiple AQIs in the package, it was decided only to include a single one to reduce the package's bloat and help ensure easy comparisons between outputs produced by the package.

The UK DAQI is based on O$_3$, NO$_2$, SO$_2$, PM$_{10}$, and PM$_{2.5}$ air pollutant concentrations. The index is split into ten categories, each further categorised into four main bands, with the thresholds seen in Table \ref{tab:ukDAQI}. The DAQI is an aggregate index, meaning that a single value is provided. The single value for the DAQI is calculated based on the highest of the set of sub-indexes based on each of the five air pollutants covered.

\begin{table}[ht]
\caption{\textbf{UK Daily Air Quality Index (DAQI) \cite{UKAIR:2023:DAQI}.} The five air pollutants comprising the Daily Air Quality Index (DAQI) are based on average concentrations across different time periods. NO$_2$ is based on the hourly mean concentrations. O$_3$ is based on the running 8-hourly mean. PM$_{10}$ is based on the 24-hour running mean. PM$_{2.5}$ is based on the 24-hour running mean. SO$_2$ is based on the 15-minute mean concentration. The colours of each column relate to the specified colour of the given DAQI level. Of note is that in its current form, to align with the data available, the DAQI calculated in the package is based upon the hourly resolution of data to simplifier the interpretation of the AQI produced rather than averaging the concentrations over their original periods, e.g. averaging hourly PM$_{10}$ over 24-hours. A benefit of this modification is allowing for AQI changes during the day to be analysed, highlighting the impact of rush hour, and making interpretations of comparisons between air pollutants much easier.}
\resizebox{0.85\textwidth}{!}{%
\begin{tabularx}{\textwidth}
{p{2.25cm}|>
{\columncolor{AQI_1}}p{0.75cm}>{\columncolor{AQI_2}}p{1.2cm}>{\columncolor{AQI_3}}p{1.3cm}>{\columncolor{AQI_4}}p{1.5cm}>{\columncolor{AQI_5}}p{1.5cm}>{\columncolor{AQI_6}}p{1.5cm}>{\columncolor{AQI_7}}p{1.5cm}>{\columncolor{AQI_8}}p{1.5cm}>{\columncolor{AQI_9}}p{1.5cm}>{\columncolor{AQI_10}}p{1.7cm}}
\textbf{Index} & \textbf{1} & \textbf{2} & \textbf{3} & \textbf{4} & \textbf{5} & \textbf{6} & \textbf{7} & \textcolor{white}{\textbf{8}} & \textcolor{white}{\textbf{9}} & \textcolor{white}{\textbf{10}}\\
\textbf{Band} & \textbf{Low} & \textbf{Low} & \textbf{Low} & \textbf{Moderate} & \textbf{Moderate} & \textbf{Moderate} & \textbf{High} & \textcolor{white}{\textbf{High}} & \textcolor{white}{\textbf{High}} & \textcolor{white}{\textbf{Very High}} \\
\cmidrule{0-10}
NO$_2$ (\si{\micro\gram/\meter^3}) & 0-67 & 68-134 & 135-200 & 201-267 & 268-334 & 335-400 & 401-467 & \textcolor{white}{468-534} & \textcolor{white}{535-600} & \textcolor{white}{601+}\\
O$_3$ (\si{\micro\gram/\meter^3}) & 0-33 & 34-66 & 67-100 & 101-120 & 121-140 & 141-160 & 161-187 & \textcolor{white}{188-213} & \textcolor{white}{214-240} & \textcolor{white}{241+}\\
PM$_{10}$ (\si{\micro\gram/\meter^3}) & 0-16 & 17-33 & 34-50 & 51-58 & 59-66 & 67-75 & 76-83 & \textcolor{white}{84-91} & \textcolor{white}{92-100} & \textcolor{white}{101+}\\
PM$_{2.5}$ (\si{\micro\gram/\meter^3}) & 0-11 & 12-23 & 24-35 & 36-41 & 42-47 & 48-53 & 54-58 & \textcolor{white}{59-64} & \textcolor{white}{65-70} & \textcolor{white}{71+}\\
SO$_2$ (\si{\micro\gram/\meter^3}) & 0-88 & 89-177 &178-266 & 267-354 & 355-443 & 444-532 & 533-710 & \textcolor{white}{711-887} & \textcolor{white}{888-1064} & \textcolor{white}{1065+}\\
\end{tabularx}
}
\label{tab:ukDAQI}
\end{table}

\noindent\textbf{Air Quality Index Functions}
\begin{itemize}
    \item Function: \textit{air\_pollution\_concentrations\_to\_UK\_daily\_air\_quality\_index} Add onto an existing dataframe the Daily Air Quality Index for the air pollutant concentration data described.
    \item Function: \textit{visualise\_air\_pollution\_daily\_air\_quality\_index} Visualise the UK Daily Air Quality Index using the individual index bounds and standard color codes.
    \item Function: \textit{visualise\_air\_pollution\_daily\_air\_quality\_bands} Visualise the UK Daily Air Quality Index using the bands and standard color codes.
\end{itemize}

\begin{figure}[!htb]
\hspace*{\fill}   % maximize separation between the subfigures
  \begin{subfigure}{0.18\textwidth}
    \includegraphics[width=\linewidth]{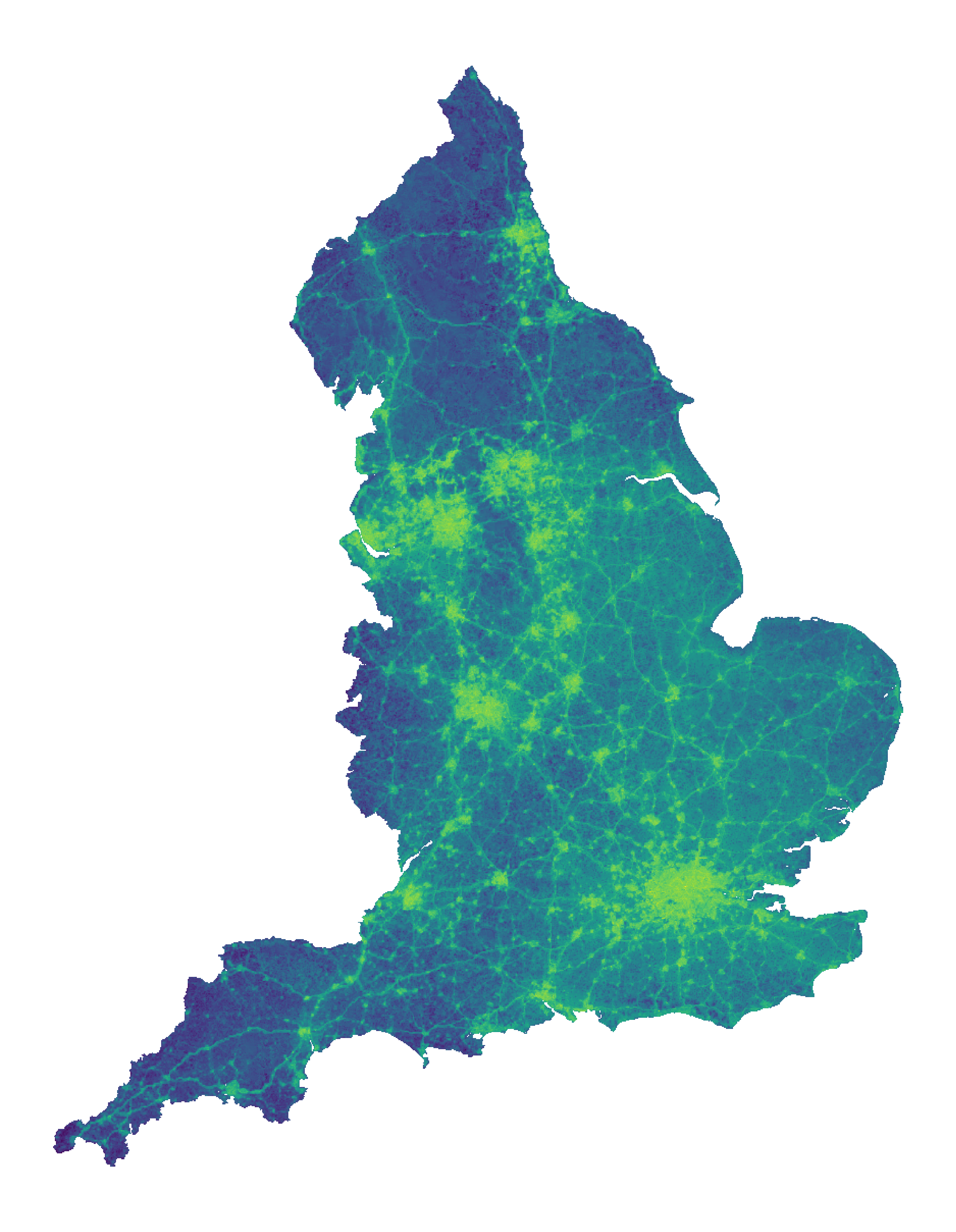}
    \caption{NO$_2$}\label{fig:exampleAQINO2}
  \end{subfigure}
  \hspace*{\fill}   % maximize separation between the subfigures
  \begin{subfigure}{0.18\textwidth}
    \includegraphics[width=\linewidth]{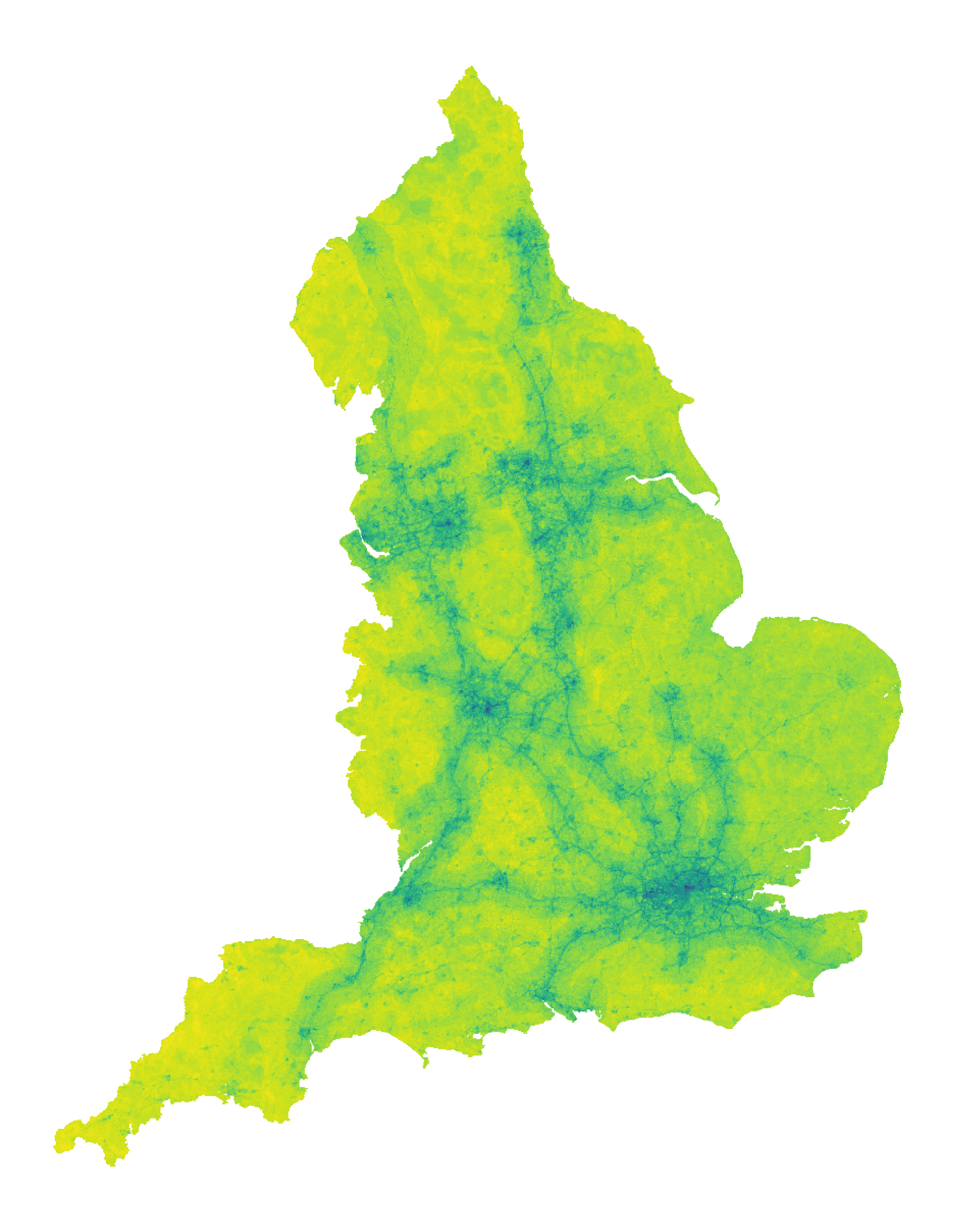}
    \caption{O$_3$}\label{fig:exampleAQIO3}
  \end{subfigure}
  \hspace*{\fill}   % maximize separation between the 
  \begin{subfigure}{0.18\textwidth}
    \includegraphics[width=\linewidth]{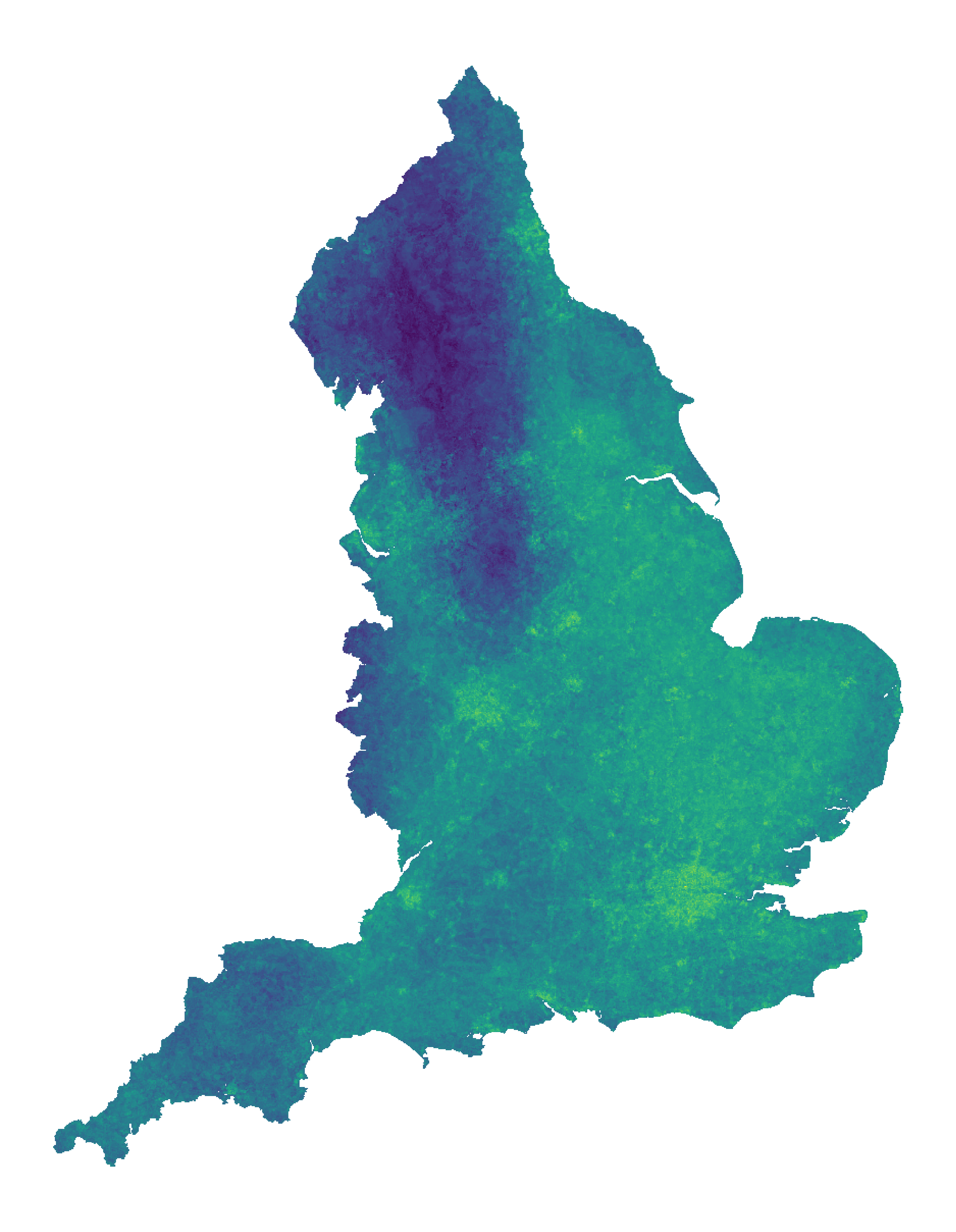}
    \caption{PM$_{10}$}\label{fig:exampleAQIPM10}
  \end{subfigure}
  \hspace*{\fill}   % maximize separation between the subfiguressubfigures
  \begin{subfigure}{0.18\textwidth}
    \includegraphics[width=\linewidth]{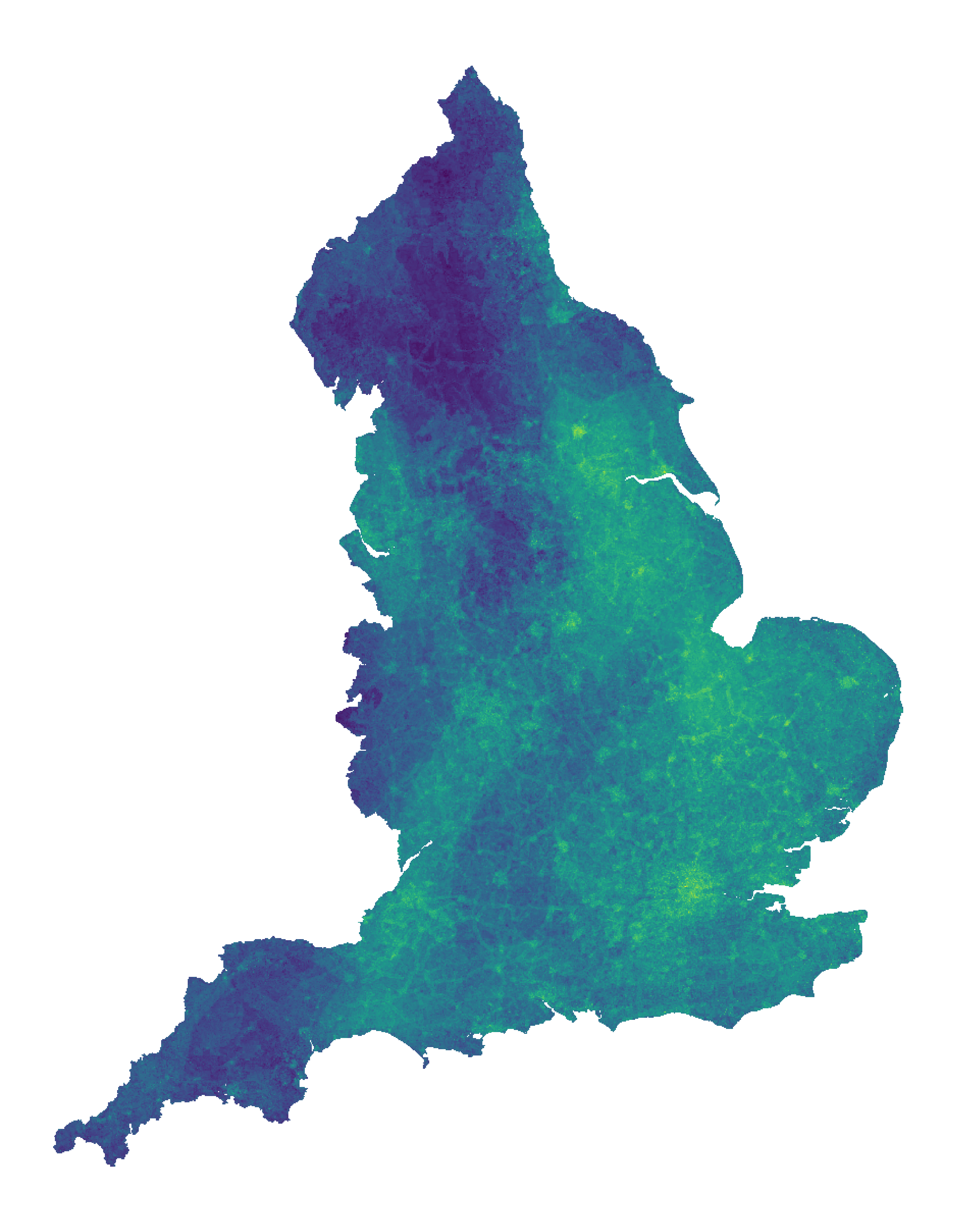}
    \caption{PM$_{2.5}$}\label{fig:exampleAQIPM25}
  \end{subfigure}
  \hspace*{\fill}   % maximize separation between the subfigures
  \begin{subfigure}{0.18\textwidth}
    \includegraphics[width=\linewidth]{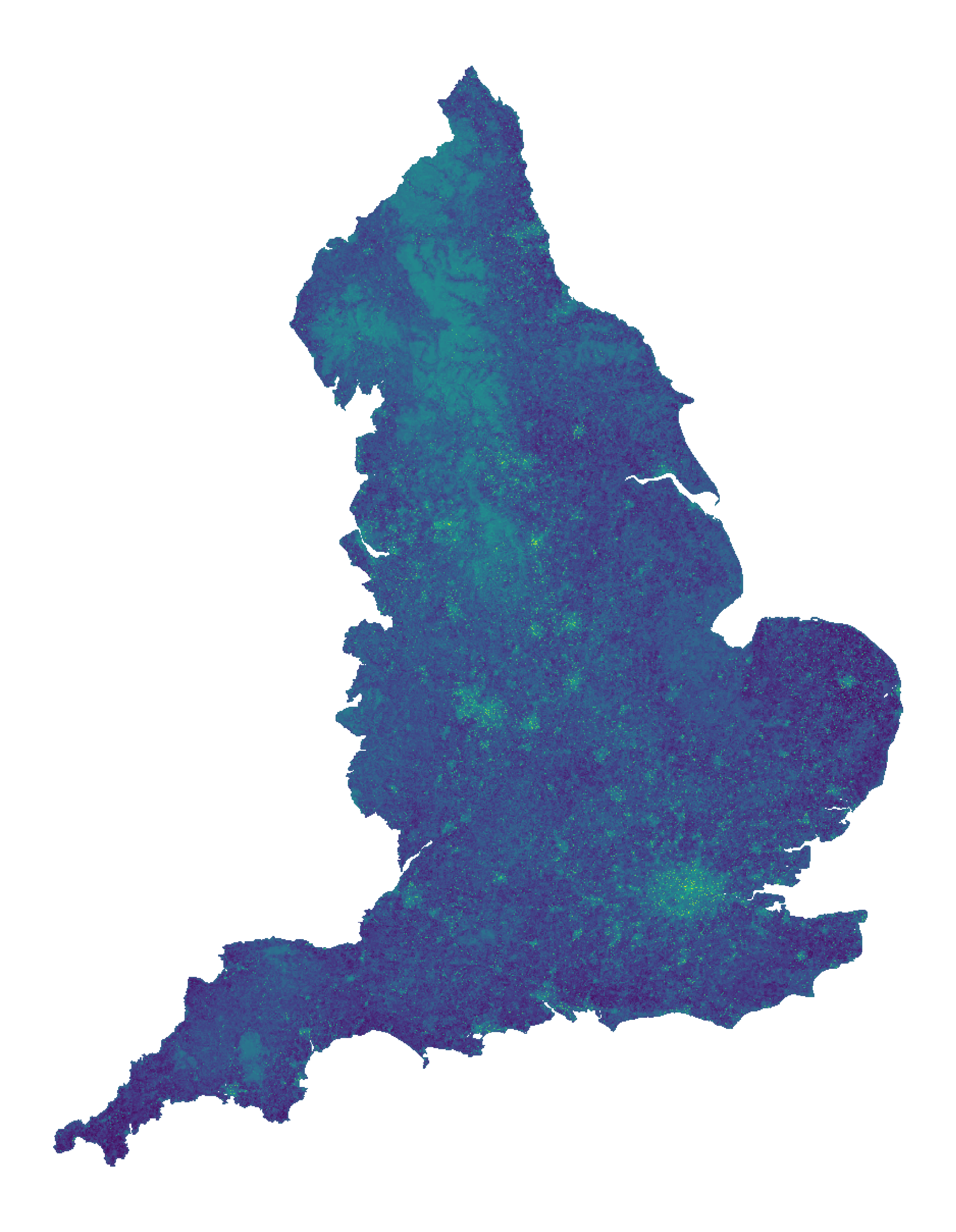}
    \caption{SO$_2$}\label{fig:exampleAQISO2}
  \end{subfigure}
  \hspace*{\fill}   % maximize separation between the 
  \\
  \hspace*{\fill}   % maximize separation between the subfigures
  \begin{subfigure}{0.3\textwidth}
    \includegraphics[width=\linewidth]{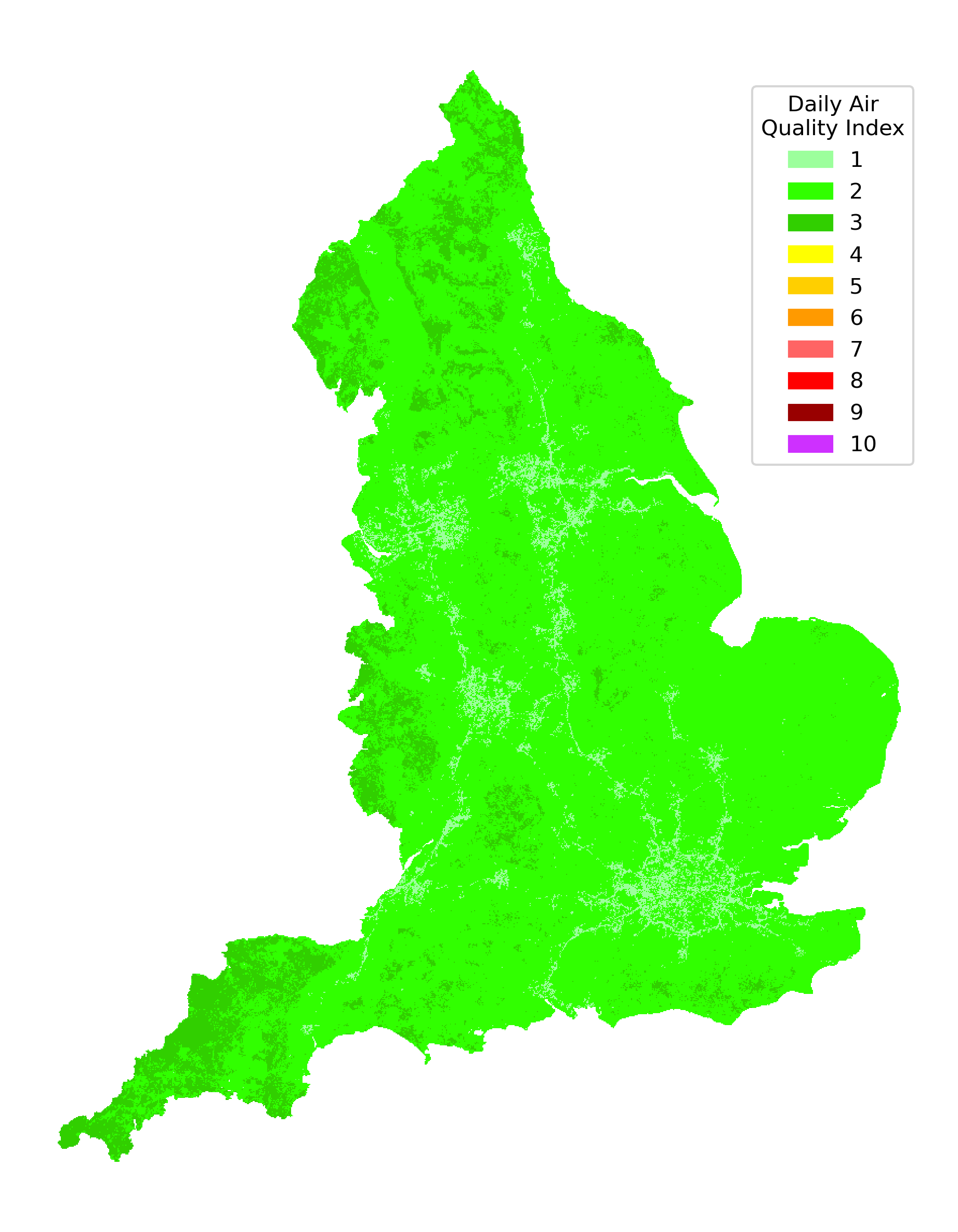}
    \caption{DAQI Index}\label{fig:exampleAQIIndex}
  \end{subfigure}
  \hspace*{\fill}   % maximize separation between the subfigures
  \begin{subfigure}{0.3\textwidth}
    \includegraphics[width=\linewidth]{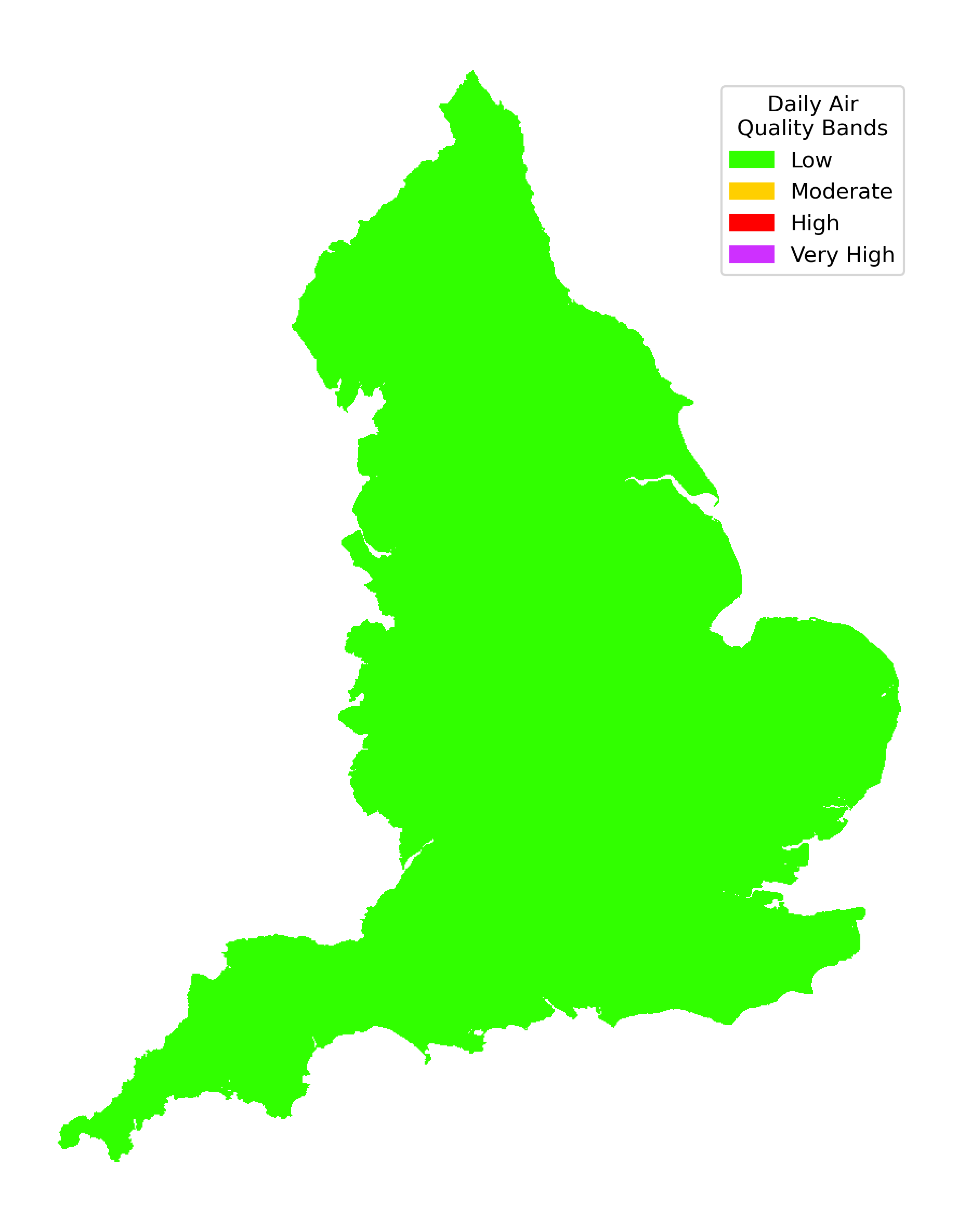}
    \caption{DAQI Band}\label{fig:exampleAQIBand}
  \end{subfigure}
  \hspace*{\fill}   % maximize separation between the subfigures
  \begin{subfigure}{0.32\textwidth}
    \includegraphics[width=\linewidth]{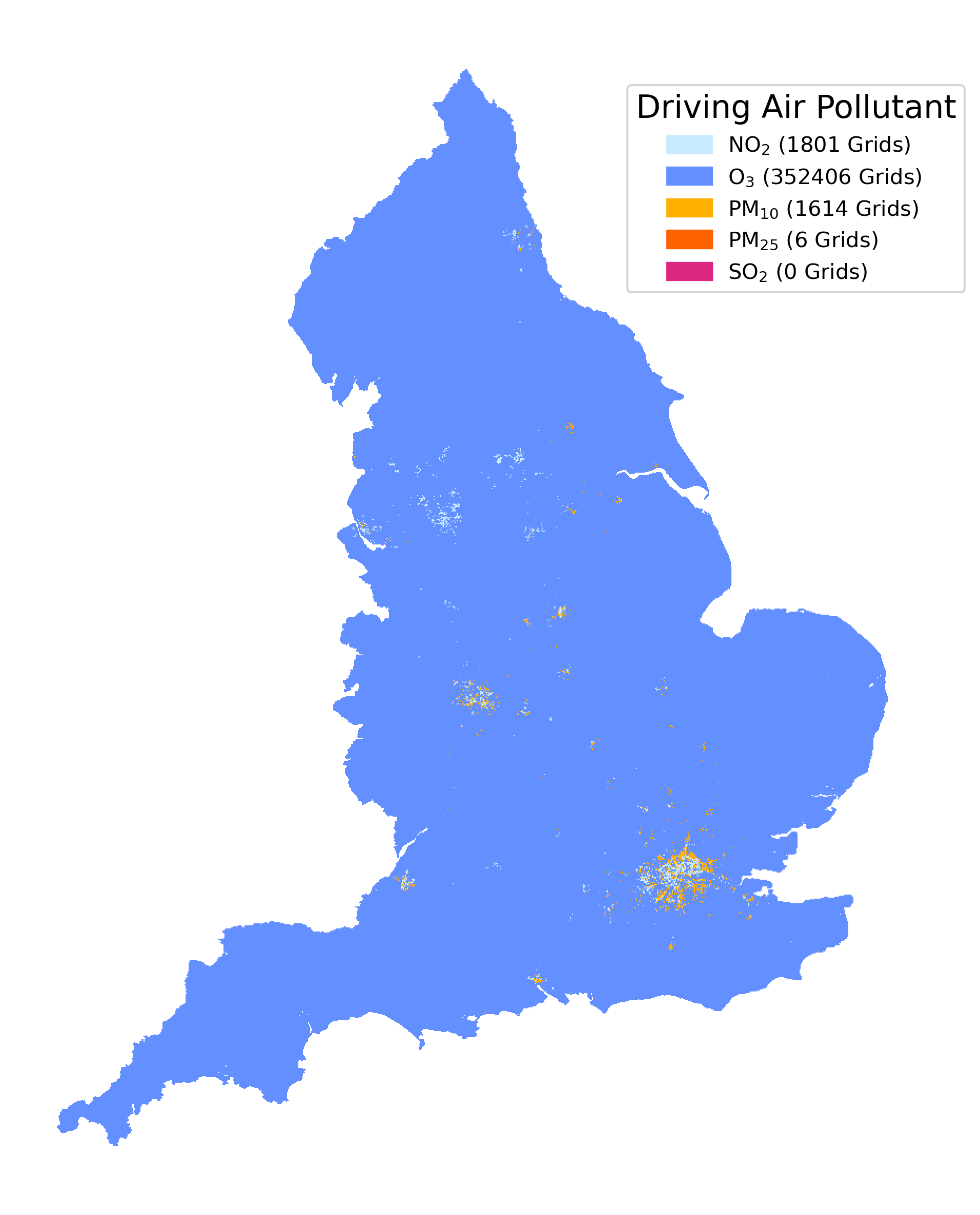}
    \caption{Driving Air Pollutant}\label{fig:drivingAirPollutants}
  \end{subfigure}
  \hspace*{\fill}   % maximize separation between the 
\caption{{\bfseries DAQI sums across all grids of the typical days of air pollution.} Each of the five air pollutants NO$_2$, O$_3$, PM$_{10}$, PM$_{2.5}$, SO$_2$ each of an assignment of a given index or band for each location, with the highest of the five determining the overall DAQI for that locations. Figure \ref{fig:drivingAirPollutants} shows how, for rural locations, O$_3$ is driving the air quality, whereas in the urban regions, particularly London, the relationship is more complex.} \label{fig:aqiBandExample}
\end{figure}

\section{Extensions To Support Stakeholders}
\subsection{A Typical Day}
\label{sec:typicalDay}

The amount of air pollution data generated at high resolutions can be overwhelming. As such, we have provided a reduced dataset alongside the original dataset for the England, showing the diurnal cycle, or a ``typical day'', for each of the locations for each day of the week for each month to try and provide a dataset that offers a general picture of air pollution concentrations, making analysis more accessible and manageable.

Rather than focusing on the actual air pollution measurements on a specific day, such as Monday, 1st January 2018, we introduce the concept of using a typical air pollution day as the baseline to explore air pollution futures. These typical air pollution days were created by averaging the conditions experienced across all common timestamps over the years the study was conducted, namely 2014-2018. The common timestamps used were a month, day of the week, and hour. This means that a common air pollution day for Monday at 8AM in January was identified and averaged all the timestamps for which this applied, such as 2014-01-06 08:00:00, 2014-01-13 08:00:00, to 2018-01-22 08:00:00, 2018-01-29 08:00:00, with 22 timestamps being used in this case over the years. The advantage of this approach was that, rather than presenting 43,824 timestamps between 2014 and 2018, only 2,016 timestamps are presented.

An added benefit of this approach is the reduction in focusing on the potentially spurious conditions experienced during a specific time. This impact allows policymakers to design interventions based on a typical day rather than a particular day and analyse the impact in a synthetic environment. This raises questions such as, "What is the impact of removing HGVs from the city centre of Exeter?" Using typical days removes potentially erroneous results from intervention designs that could otherwise be focused on days exhibiting atypical conditions.

\noindent\textbf{Typical Day Function}
\begin{itemize}
    \item Function \textit{air\_pollution\_concentration\_typical\_day\_real\_time\_united\_kingdom} Retrieve the typical day complete dataset for England for a given time.
    \item Function \textit{air\_pollution\_concentration\_nearest\_point\_typical\_day\_united\_kingdom} Retrieve a single air pollution concentration data point predicted based on the England data, based on the cloest point given by the latitude and longitude.
\end{itemize}

\begin{figure}[!htb]
  \hspace*{\fill}   % maximize separation between the subfigures
  \begin{subfigure}{0.32\textwidth}
    \includegraphics[width=\linewidth]{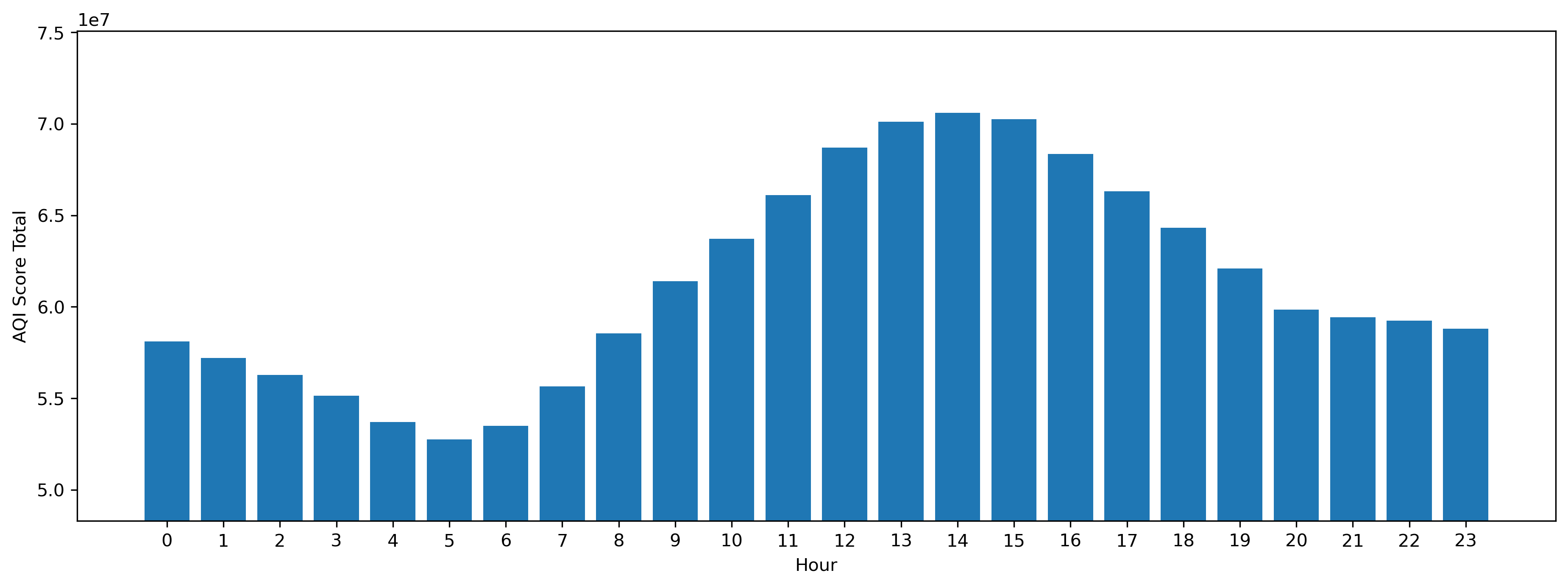}
    \caption{AQI Total Hour}\label{fig:aqiTotalHour}
  \end{subfigure}
  \hspace*{\fill}   % maximize separation between the subfigures
  \begin{subfigure}{0.32\textwidth}
    \includegraphics[width=\linewidth]{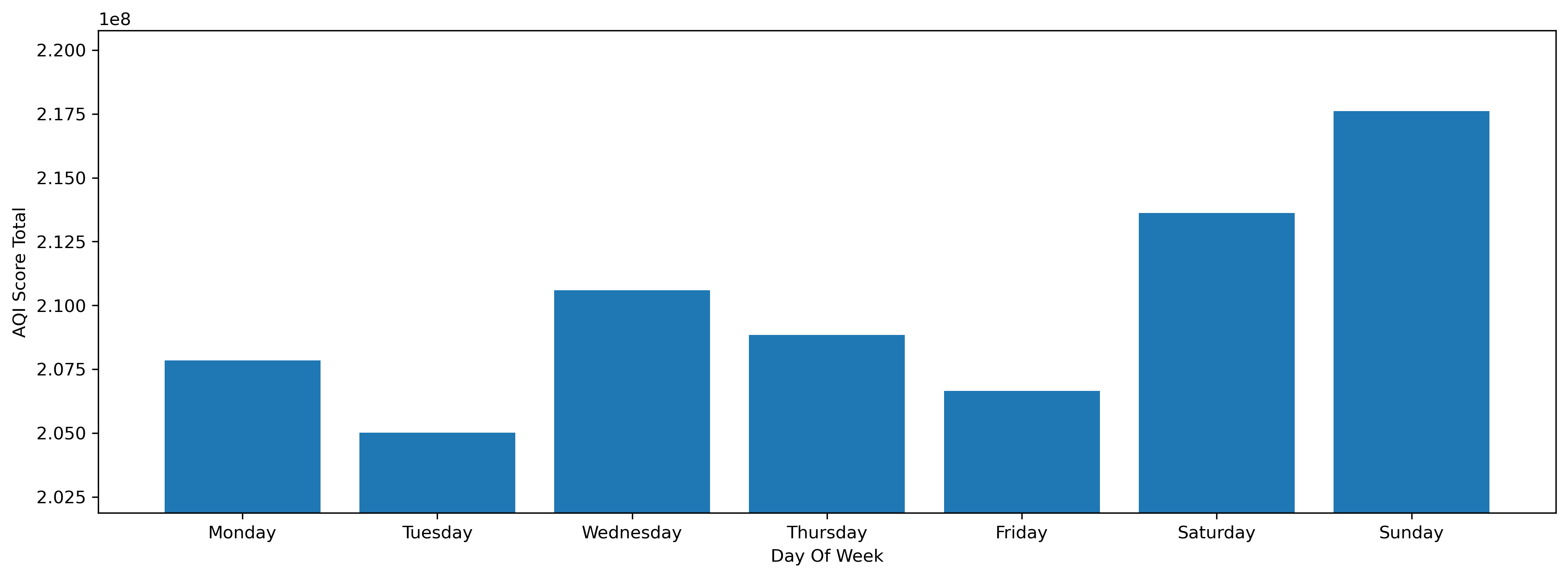}
    \caption{AQI Total Day}\label{fig:aqiTotalDay}
  \end{subfigure}
  \hspace*{\fill}   % maximize separation between the subfigures
  \begin{subfigure}{0.32\textwidth}
    \includegraphics[width=\linewidth]{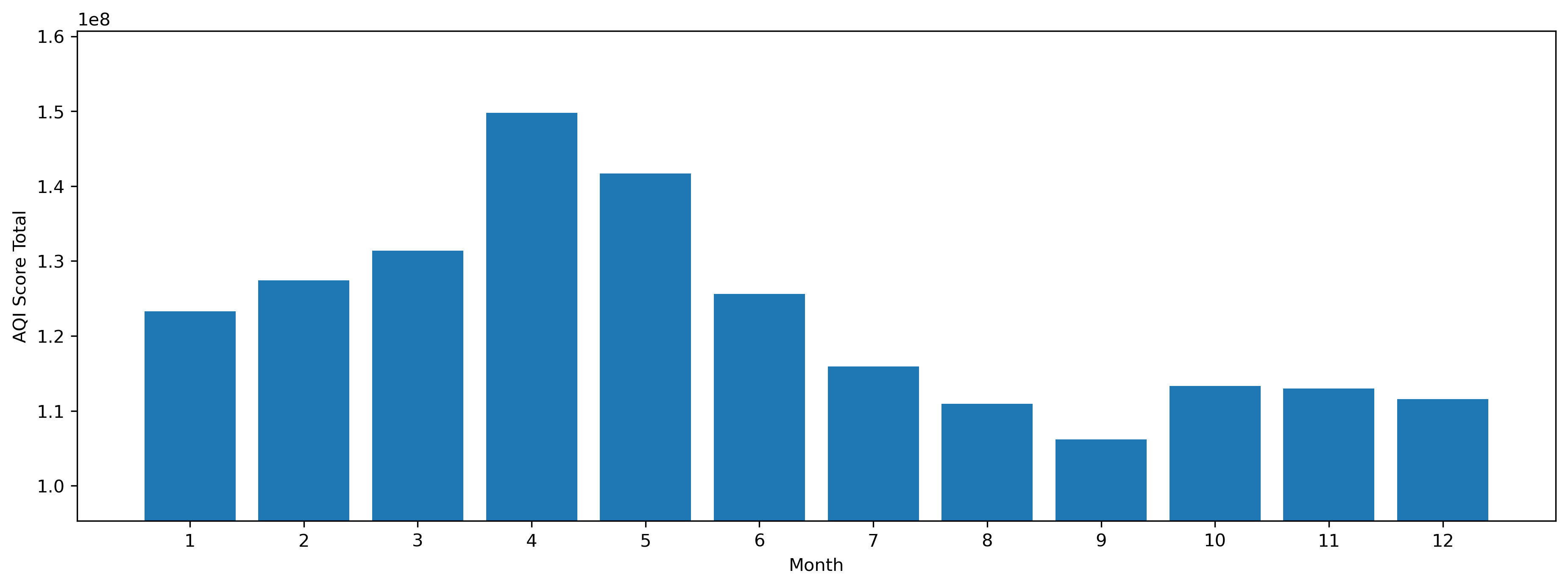}
    \caption{AQI Total Month}\label{fig:aqiTotalMonth}
  \end{subfigure}
  \hspace*{\fill}   % maximize separation between the subfigures
\caption{{\bfseries DAQI sums across all grids of the typical days of air pollution.} Similarly to Figure \ref{fig:aqiBandExample}, where spatially different air pollutants drive poor air quality, the same can be true temporally. For example, the high AQI sum on Sunday is driven by O$_3$, a well-observed phenomena \cite{sicard:2020:OzoneWeekend}. In contrast, the peaks in the later parts of each day are likely more complex, driven by potentially NO$_2$ concentrations accumulating from rush hour or O$_3$ production from a day of intense sunshine.} \label{fig:aqiTotalComparison}
\end{figure}

\subsection{Prediction Intervals}
\label{sec:predictionInterval}

\begin{figure}[!htb]
  \hspace*{\fill}   % maximize separation between the subfigures
  \begin{subfigure}{0.25\textwidth}
    \includegraphics[width=\linewidth]{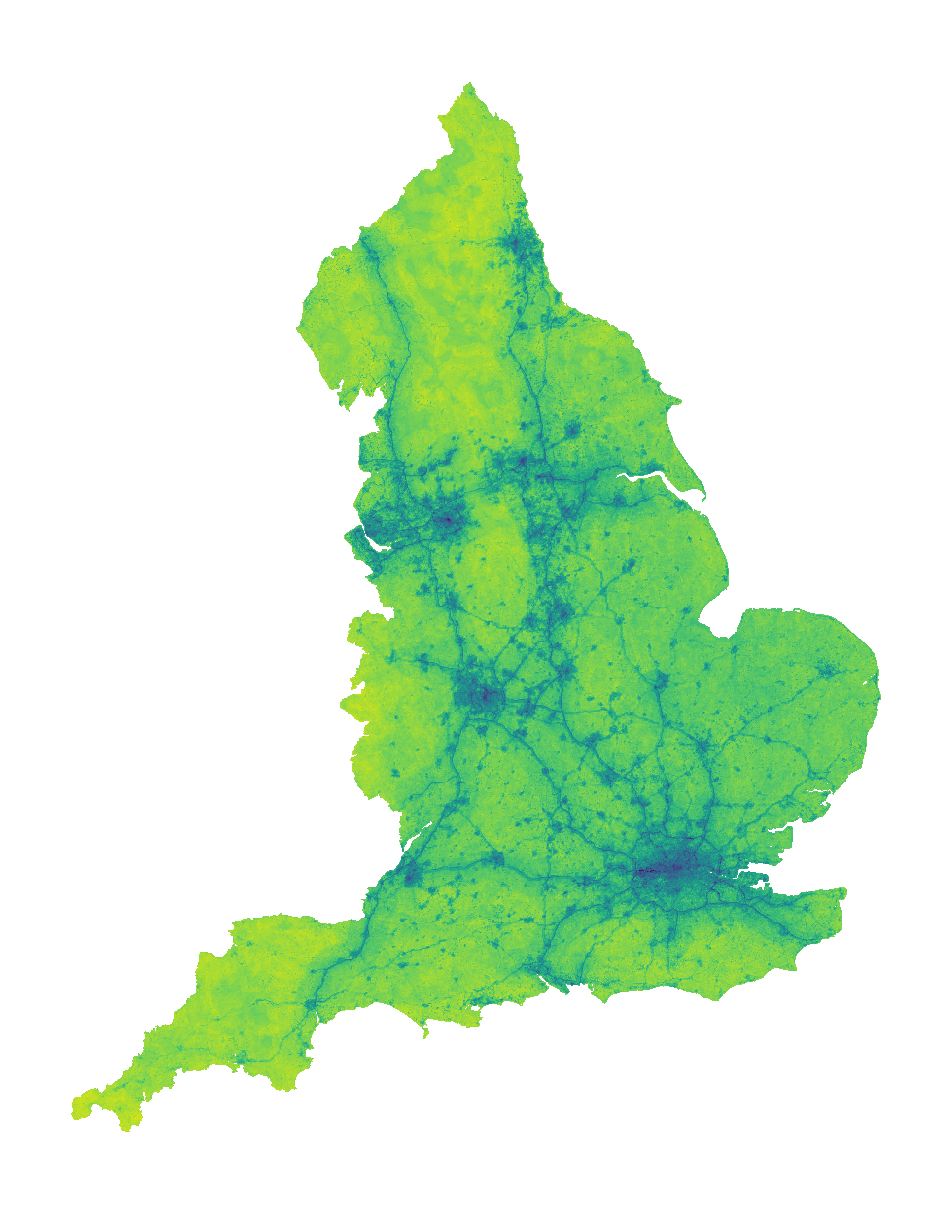}
    \caption{Lower Bound, 0.05 Quantile}\label{fig:fullSpatialPollutionMapPredicitionIntervalNOLowerBound}
  \end{subfigure}%
  \hspace*{\fill}   % maximize separation between the subfigures
  \hspace*{\fill}   % maximize separation between the subfigures
  \begin{subfigure}{0.25\textwidth}
    \includegraphics[width=\linewidth]{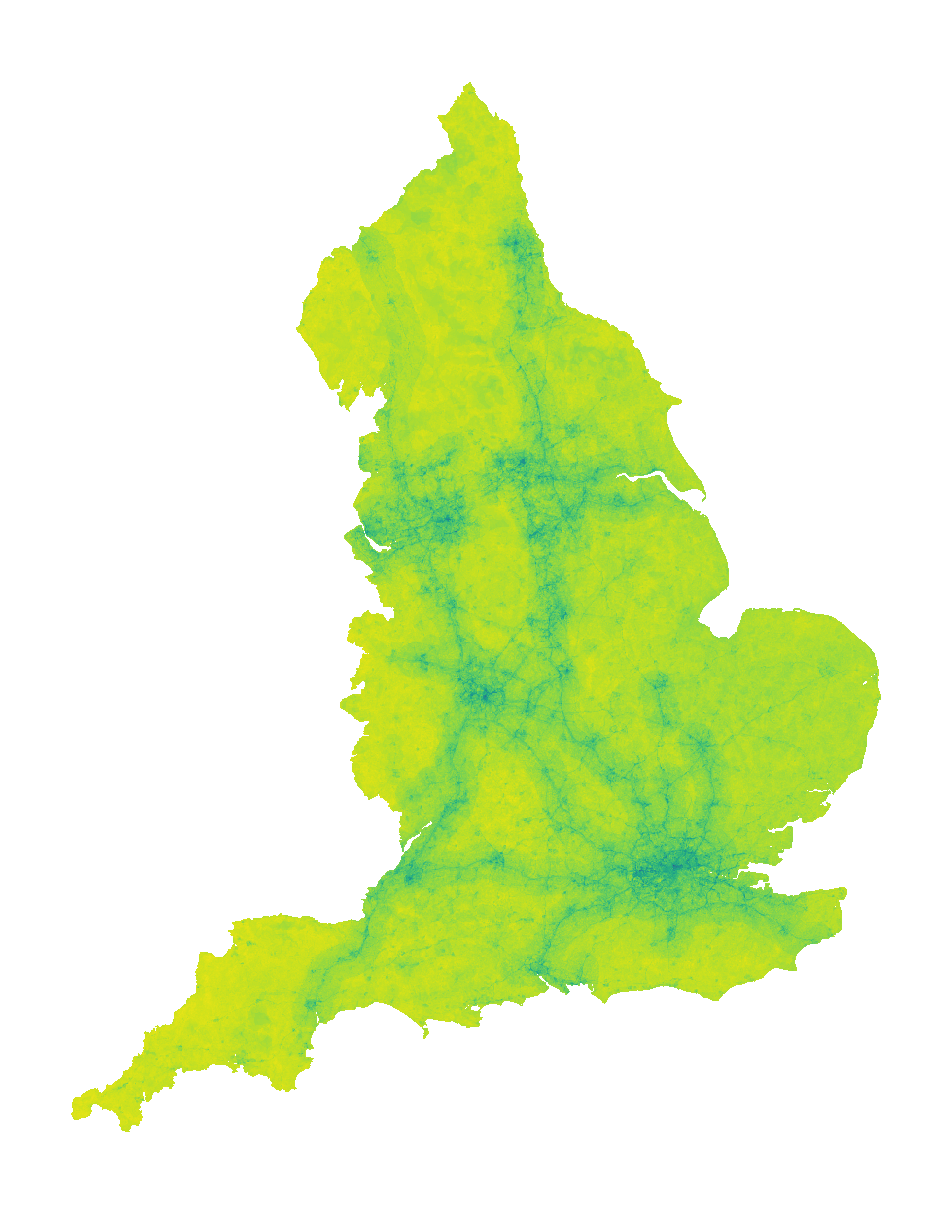}
    \caption{O$_{3}$ Median, 0.5 Quantile} \label{fig:fullSpatialPollutionMapPredicitionIntervalNO2Median}
  \end{subfigure}%
  \hspace*{\fill}   % maximize separation between the subfigures
  \hspace*{\fill}   % maximize separation between the subfigures
  \begin{subfigure}{0.25\textwidth}
    \includegraphics[width=\linewidth]{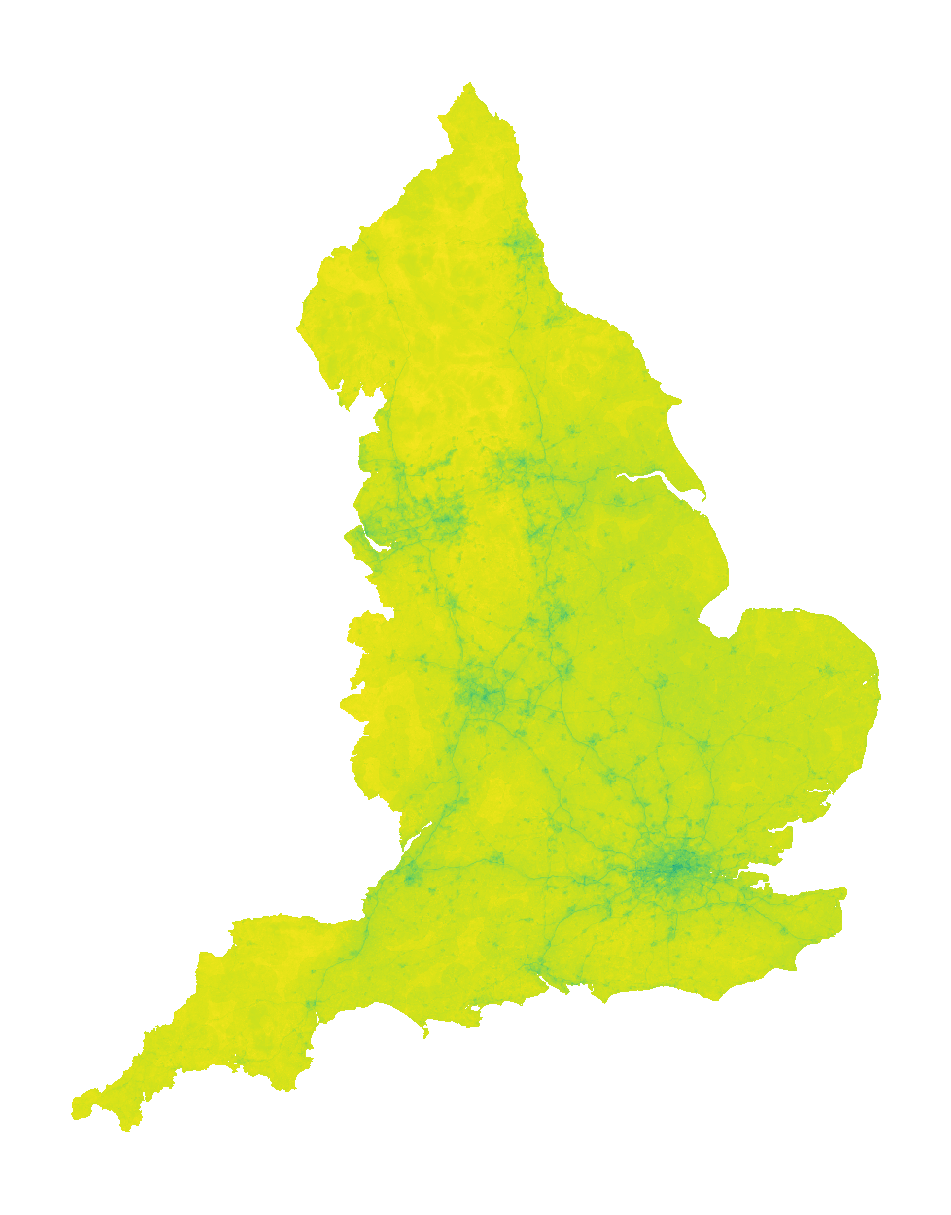}
    \caption{Upper Bound, 0.95 Quantile}\label{fig:fullSpatialPollutionMapPredicitionIntervalNO2UpperBound}
  \end{subfigure}%
  \raisebox{20mm}{
  \begin{subfigure}{0.12\textwidth}
    \includegraphics[width=\linewidth]{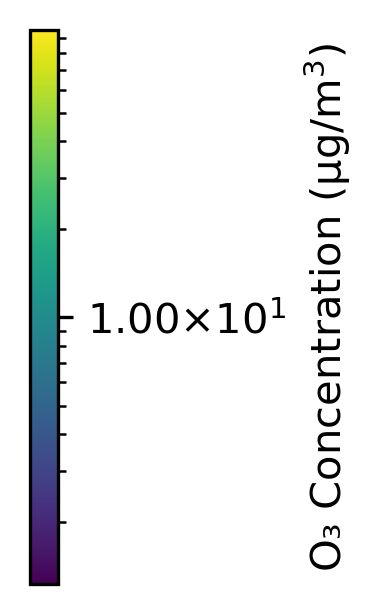}
  \end{subfigure}}
  \\
  \hspace*{\fill}   % maximize separation between the subfigures
  \begin{subfigure}{0.5\textwidth}
    \includegraphics[width=\linewidth]{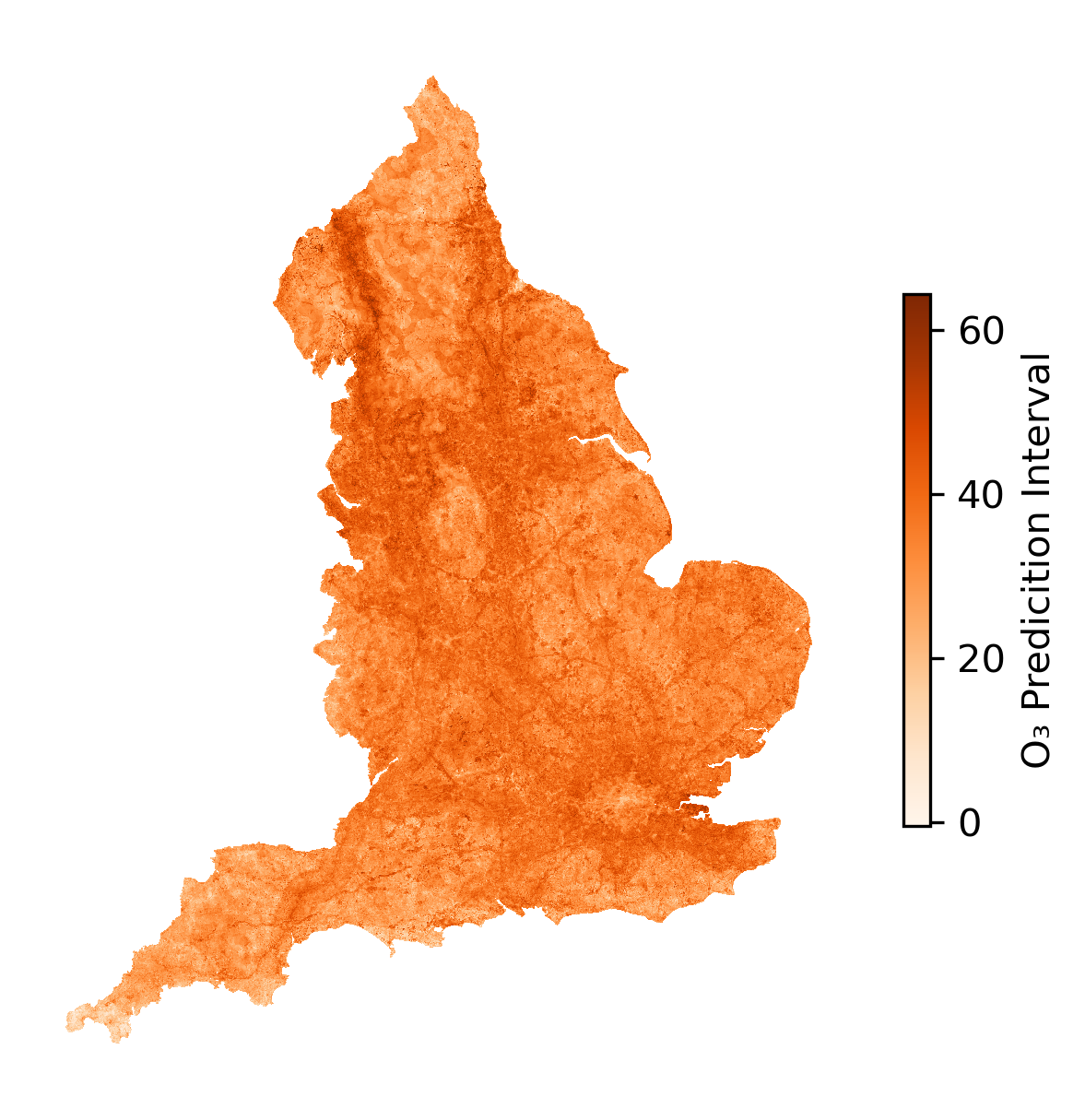}
     \caption{90\% Prediction Interval Size}
  \end{subfigure}%
  \hspace*{\fill}   % maximize separation between the subfigures
\caption{{\bfseries Prediction Interval examples for full spatial map of England.} Further to individual locations, the prediction interval provides insight into the locations in which the model is less sure about predictions, with the size of the interval denoting a more extensive range of possible values a prediction can take.} \label{fig:fullSpatialPollutionMapPredicitionInterval}
\end{figure}

The output of the wide range of air pollution models discussed in Section \ref{sec:existingSoftware} is a single value for the air pollution concentration. The limitation of this approach is its lack of insight into the model's confidence in the estimation made. Utilising the same model framework as in previous work \cite{berrisford:2024:EstimatingAirPollutionHourlyGlobal}, it is feasible to employ quantile regression, which can estimate the median or other user-defined quantiles of the target vector distribution. This framework facilitates the creation of a prediction interval by training a model at the 0.05 and 0.95 quartiles, providing a baseline estimation of the median with the prediction of the 0.5 quartile. The prediction interval quantifies the uncertainty associated with a given prediction, indicating the range of possible values for a particular prediction, thus answering the question, ``Where could a future prediction possibly fall?''. Employing the 0.05 and 0.95 quartiles offers a 90\% confidence level that a future observation, given specific predictors, will fall within the denoted interval. Including the prediction interval aids policymakers and underscores scenarios where the model output should not be used, and alternative data should be sought—such as outputs from physical-chemical models or actual air pollution concentration measurements—especially when the prediction interval exceeds the user's risk appetite. Figure \ref{fig:fullSpatialPollutionMapPredicitionInterval} illustrates the model prediction for the median and the 0.05 and 0.95 quantiles bounding the 90\% prediction interval for the grid where the Chesterfield Loundsley Green monitoring station is located, demonstrating how the interval varies at different points in the time series and highlighting instances where the model has lower confidence in its predictions. All available data via the functions outlined in Section \ref{sec:dataAirPollutionConcentrations} include the associated median and prediction interval predictions within the Python package.

\begin{figure}[!htb]
    \includegraphics[width=\linewidth]{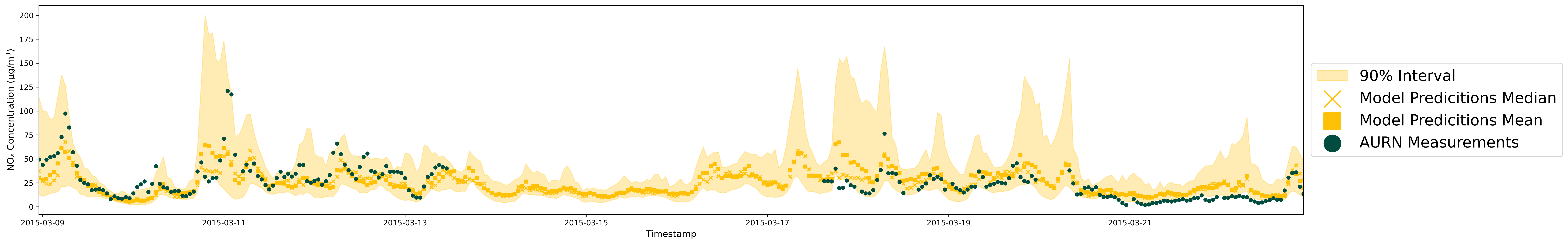}
\caption{{\bfseries Chesterfield Loundsley Green Road NO$_x$ Emissions.} The AURN measurements for the Chesterfield Loundsley Green Road monitoring station are shown alongside a range of different model predictions for air pollution concentrations for NO$_x$. Of particular note is the size of the 90\% interval, where the predictions for the model mean are broadly similar between 2015-03-15 and 2013-03-18; however, the prediction interval provides considerable further assurance of the prediction. The prediction interval highlights that the predictions for 2013-03-18 are considerably less certain.} \label{fig:aurnModelComparisonPredicitionInterval}
\end{figure}

\section{Designing Interventions}
\subsection{Accessing Models}

The key contribution of the Python package proposed in this work is the utilisation of the underlying models used to create the complete datasets alongside the feature vector data. This approach enables users to manipulate the feature vector and make predictions about potential air pollution futures. Due to the considerable size of the data (100GBs), the feature vector dataset made available is the typical data feature vectors described in Section \ref{sec:typicalDay}.

\noindent\textbf{Model and Feature Vector Accessing Functions}
\begin{itemize}
    \item Function \textit{get\_model\_feature\_vector\_names} Getter function to return the list of features that were used for a given model type.
    \item Function \textit{make\_concentration\_predicitions\_united\_kingdom} Make predicition for a given environment conditions for air pollution concentrations.
    \item Function \textit{load\_feature\_vector\_typical\_day\_united\_kingdom} Load in a feature vector for the typical day in the United Kingdom.
    \item Function \textit{load\_model\_united\_kingdom} Load in a pre trained air pollution machine learning model for England.
    \item Function \textit{load\_model\_global} Load in a pre trained air pollution machine learning model for the globe.
\end{itemize}

\subsection{Framework For Comparison}

\begin{figure}[!htb]
  \hspace*{\fill}   % maximize separation between the subfigures
  \begin{subfigure}{0.24\textwidth}
    \includegraphics[width=\linewidth]{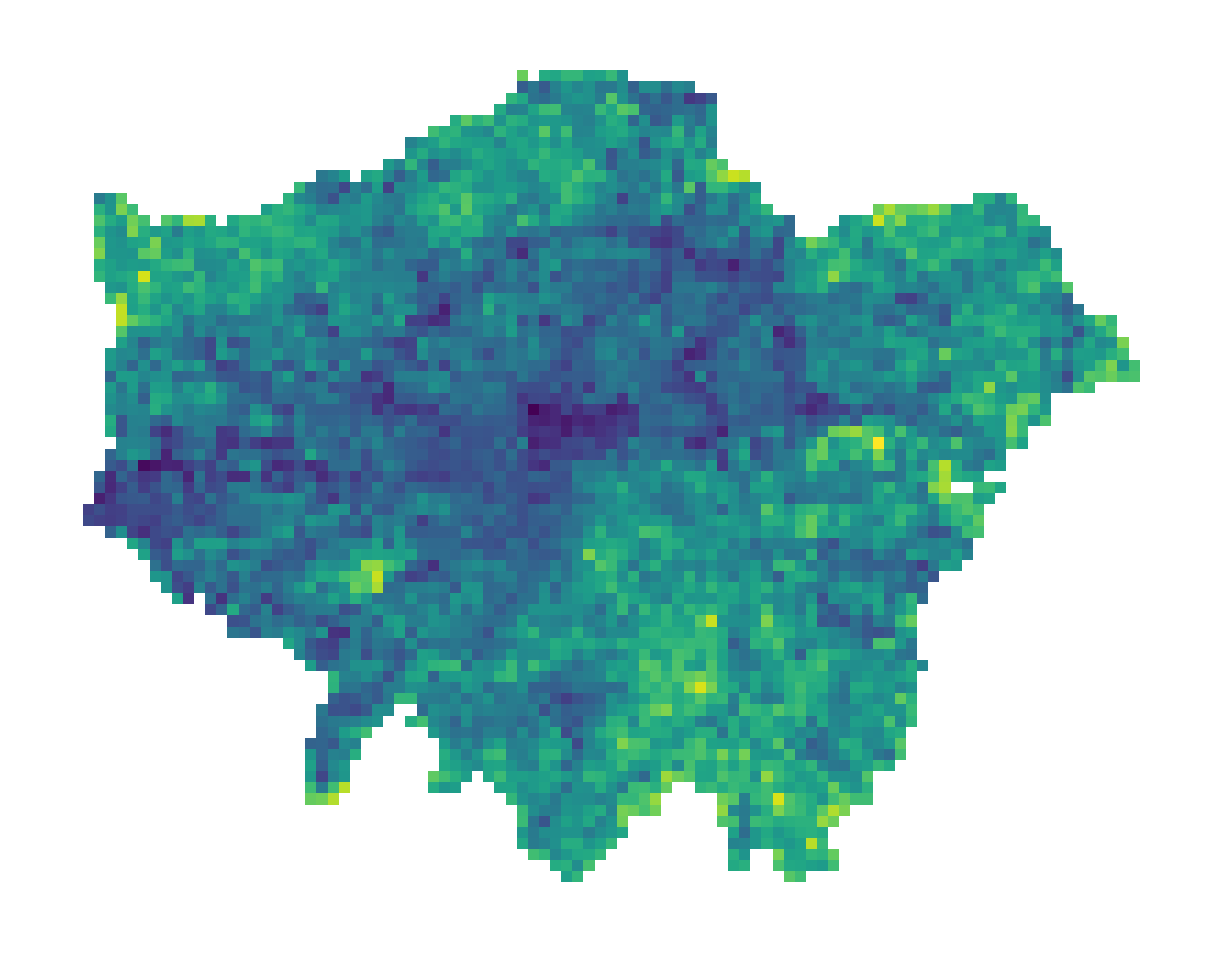}
    \caption{O$_3$ Current}\label{fig:comparisonBaseline}
  \end{subfigure}
  \hspace*{\fill}   % maximize separation between the subfigures
  \begin{subfigure}{0.24\textwidth}
    \includegraphics[width=\linewidth]{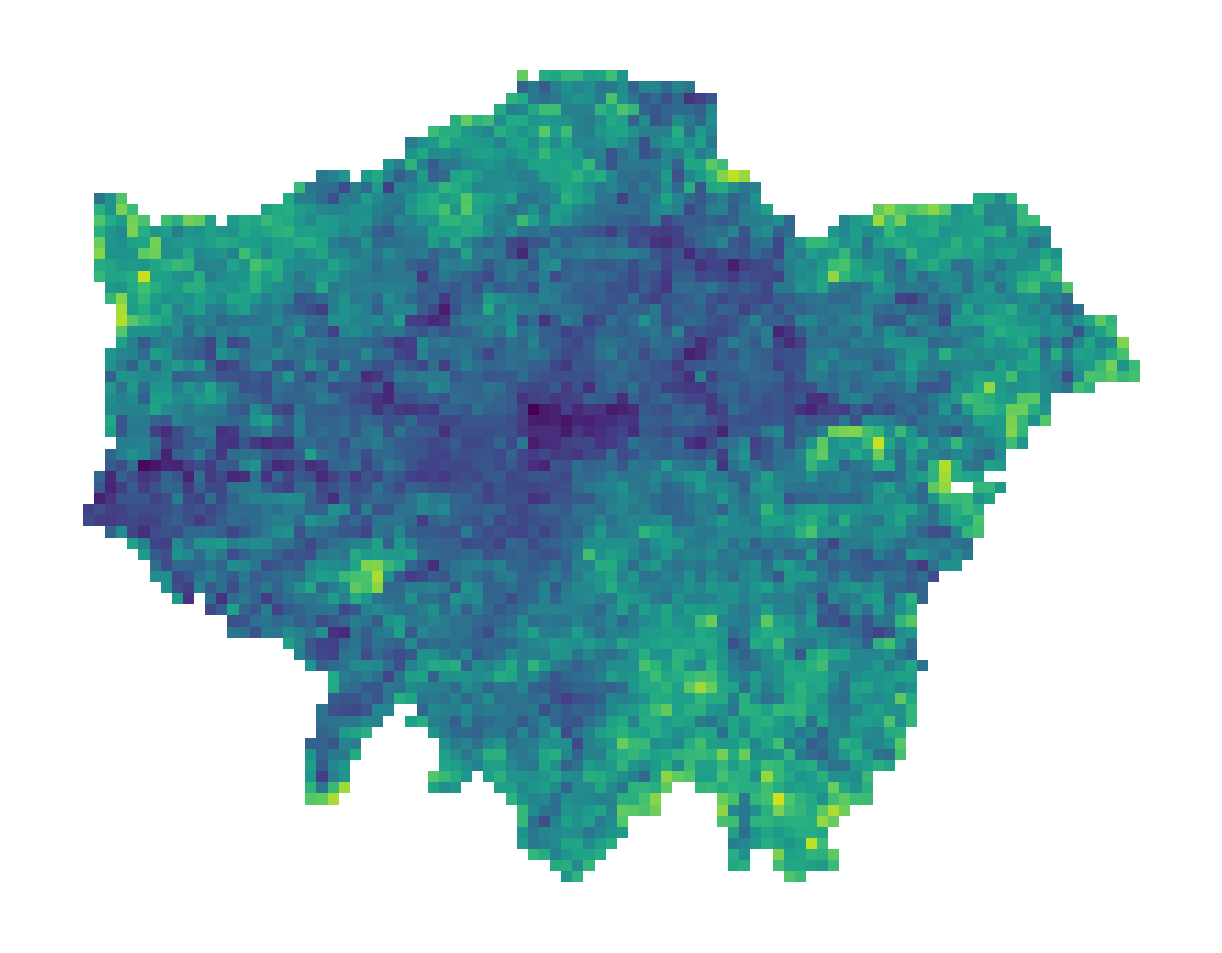}
    \caption{O$_3$ Future}\label{fig:comparisonFuture}
  \end{subfigure}
  \hspace*{\fill}   % maximize separation between the subfigures
  \raisebox{10mm}{
  \begin{subfigure}{0.06\textwidth}
    \includegraphics[width=\linewidth]{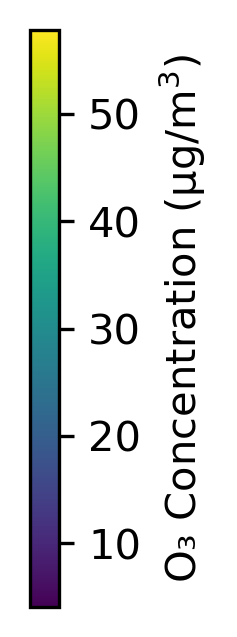}
  \end{subfigure}}
  \hspace*{\fill}   % maximize separation between the subfigures
  \begin{subfigure}{0.28\textwidth}
    \includegraphics[width=\linewidth]{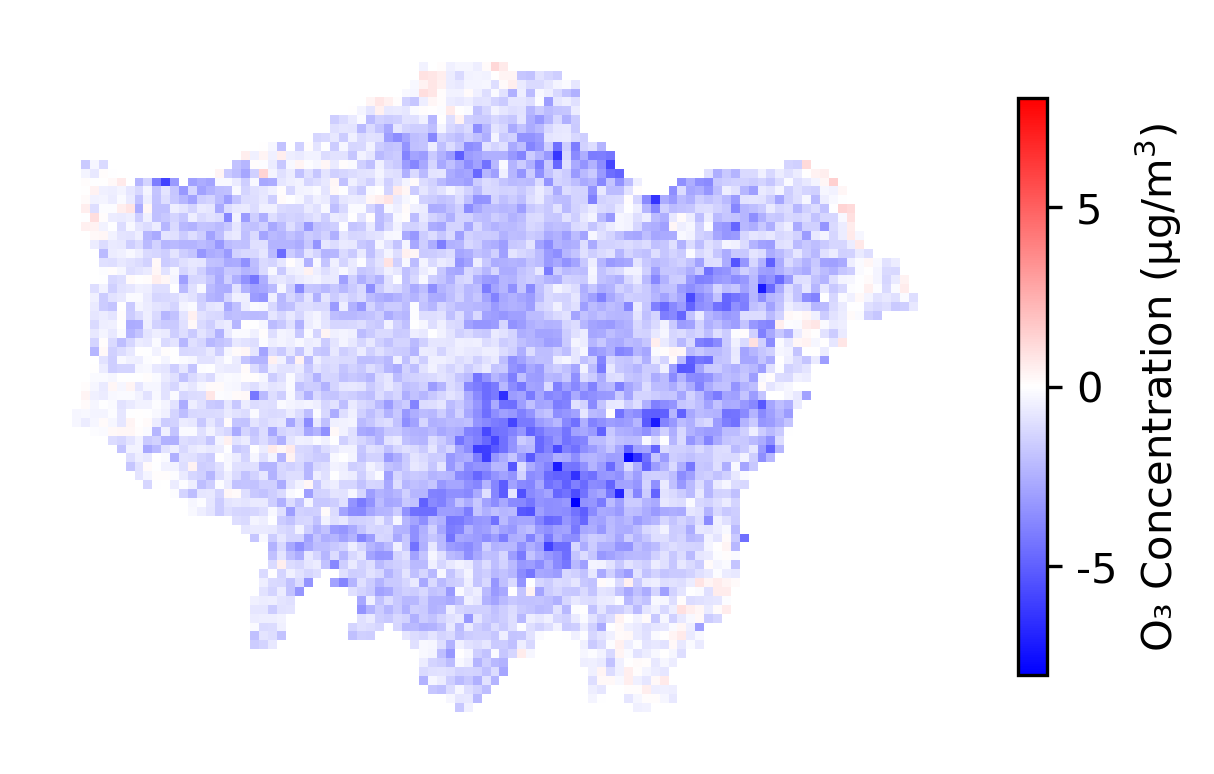}
    \caption{O$_3$ Change}\label{fig:comparisonChange}
  \end{subfigure}
  \hspace*{\fill}   % maximize separation between the subfigures
\caption{{\bfseries Comparison of future states of O$_3$ air pollution concentrations across Greater London, when decreasing wind by 20\%.} Decreasing wind speed highlights the relationship between O$_3$ and wind, assuming all other variables remain the same. The models allow for quick computation of future hypothetical scenarios depending on the user's desires, allowing for more complex scenarios, such as changing wind speed, rainfall, and transportation.} \label{fig:aqiMapComparisons}
\end{figure}

Alongside creating possible air pollution futures, the package provides functionality for making visualisations of these futures and comparisons between them, as illustrated in Figure \ref{fig:aqiMapComparisons} for the scenario of wind decreasing by 20\% on the typical day of January Monday at 8 AM across Greater London.

\noindent\textbf{Air Pollution Scenario Comparison Functions}
\begin{itemize}
    \item Function \textit{change\_in\_concentrations\_visulisation} Visualise the change in concentrations for two datasets of air pollution concentrations based on actual concentrations. 
    \item Function \textit{change\_in\_aqi\_visulisation}  Visualise the change in concentrations for two datasets of air pollution concentrations based on air quality indexes.
    \item Function \textit{change\_in\_concentration\_line}  Visualise the change in concentrations for two datasets of air pollution concentrations in a line graph. Examples can be seen in Figure \ref{fig:hardInterventionLineConcentrations}.
\end{itemize}

\section{Making Informed Decisions}

The purpose of having high-resolution spatial and temporal air pollution concentration data is to enable stakeholders to make informed decisions. We identify two types of interventions: soft and hard. A soft intervention is defined as taking action to avoid air pollution, such as an individual altering their behaviour to reduce their exposure to air pollution by changing the route of their daily commute. Conversely, a hard intervention is described as something more long-term and strategic, such as the placement of new infrastructure.

\subsection{Soft Interventions}

As an example of a possible soft intervention into air pollution exposures that can be conducted, we have examined air pollution concentrations across critical locations throughout England. The locations we investigated include schools, hospitals, and prisons, chosen because they are places where attendance is not typically voluntary but is dictated by circumstances beyond an individual's control. Given this, we consider it paramount that the air pollution levels at these locations remain within safe limits, with efforts made to further reduce exposure as much as possible, given the constraints of the locations.

\begin{figure}
\hspace*{\fill}   % maximize separation between the subfigures
\raisebox{4mm}{
  \begin{subfigure}{0.3\textwidth}
    \includegraphics[width=\linewidth]{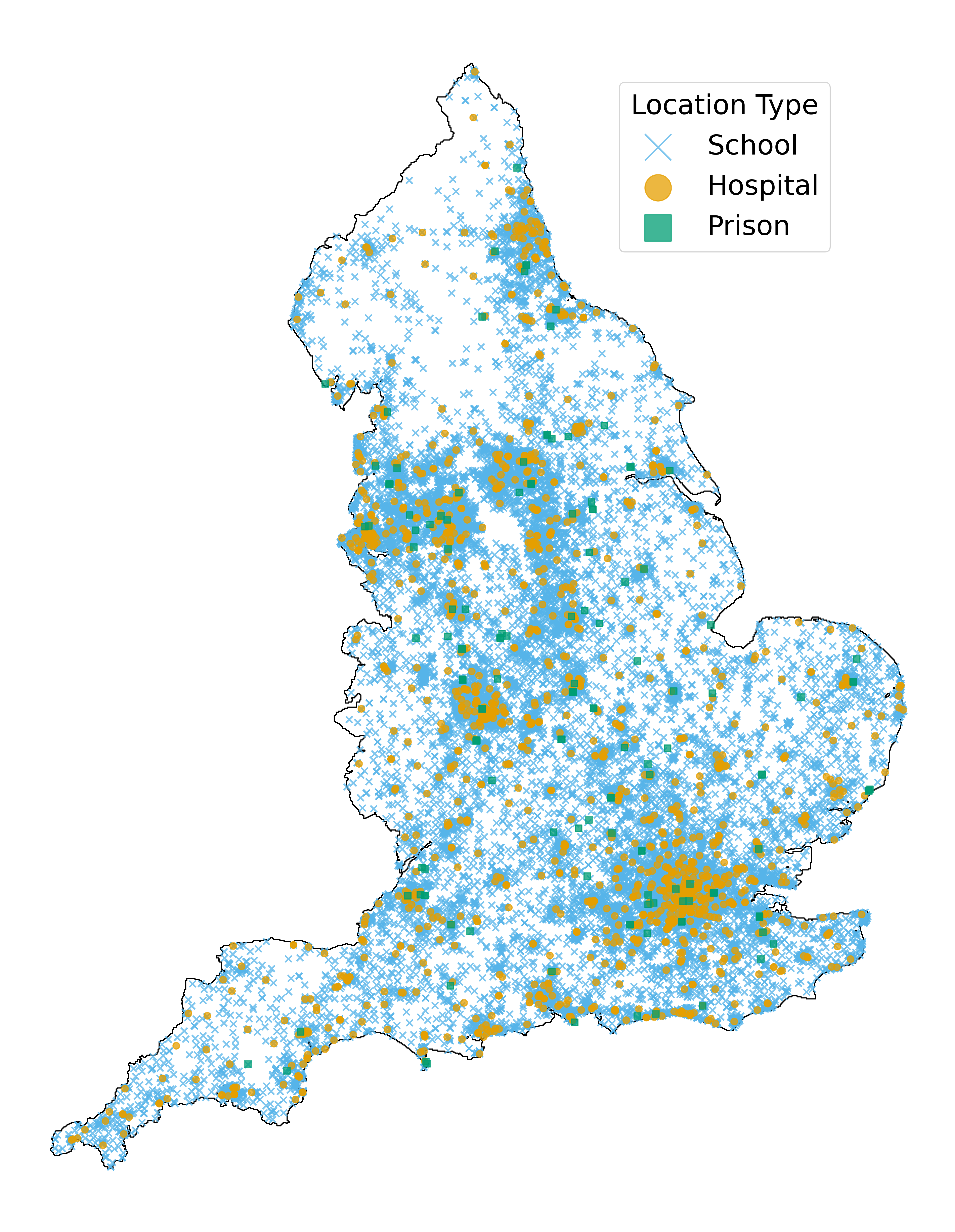}
  \end{subfigure}%
  }
  \hspace*{\fill}   % maximize separation between the subfigures
  \begin{subfigure}{0.65\textwidth}
    \includegraphics[width=\linewidth]{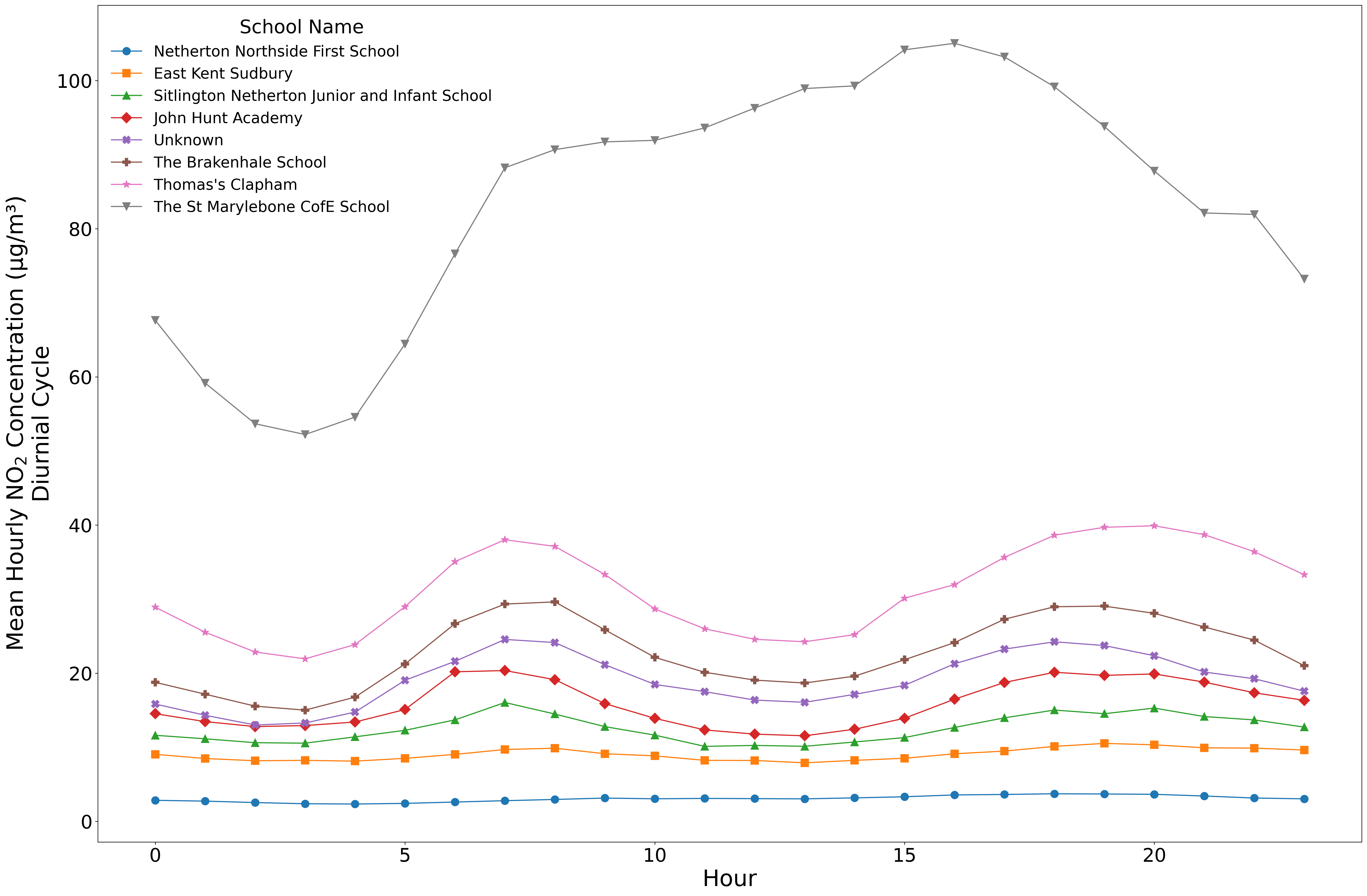}
  \end{subfigure}%
  \hspace*{\fill}   % maximize separation between the subfigures
\caption{{\bfseries Locations of key locations within England from OpenStreetMaps, alongside the air pollution diurnial cycle for a subset of schools.} The time series of the subsets of schools highlights the inequality between different locations, with more rural schools experiencing considerably better air quality. The time series also allows for proactive planning of behaviours in the context of air pollution, such as scheduling outdoor lessons during the dips in air pollution concentrations.} \label{fig:softIntervention}
\end{figure}

Figure \ref{fig:softIntervention} displays critical locations across England alongside a time series of the diurnal cycle of air pollution for a subset of schools. The figure clearly illustrates the disparity in air pollution exposure between schools, notably between rural and urban settings. Netherton Northside First School, situated in northern England, consistently experiences good air quality, in contrast to The St Marylebone CofE School in central London, which endures air pollution levels significantly higher at all times, far exceeding the WHO guideline of 10\si{\micro\gram/\meter^3} \cite{WHO:2021:GlobalAQG}.

While the data provides insight into inequality between schools, it also enables the design and implementation of simple, achievable interventions. For instance, at The St Marylebone CofE School, the air pollution levels are observed to be lower in the morning than in the afternoon. This observation opens the possibility of scheduling outdoor lessons and break times in the morning when the air quality is better, and planning more indoor activities in the afternoon.

A further note is that Figure \ref{fig:softIntervention} displays an average for the diurnal cycle, with the underlying data being unique to each hour across the dataset. Analyses similar to those shown in Figure \ref{fig:aqiTotalComparison} are possible for each location, offering insights not only into the optimal hours for outdoor lessons but also into the best days of the week and months to minimise air pollution exposure. This example directly responds to the concerns of school stakeholders, such as parents wishing to understand the air pollution levels around a given school.\footnote{\href{https://www.sundaypost.com/fp/air-quality-monitors-outside-every-primary-school/?utm_source=twitter}{School and Air Pollution. (Last Accessed: 05/03/2024)}}

\noindent\textbf{Soft Intervention Functions}
\begin{itemize}
    \item Function \textit{fetch\_amenities\_as\_geodataframe} Fetch amenities from OpenStreetMaps of a given type within a bounding box and return as a GeoDataFrame.
\end{itemize}

\subsection{Hard Interventions}

\begin{figure}
  \hspace*{\fill}   % maximize separation between the subfigures
  \raisebox{-2mm}{
  \begin{subfigure}{0.35\textwidth}
    \includegraphics[width=\linewidth]{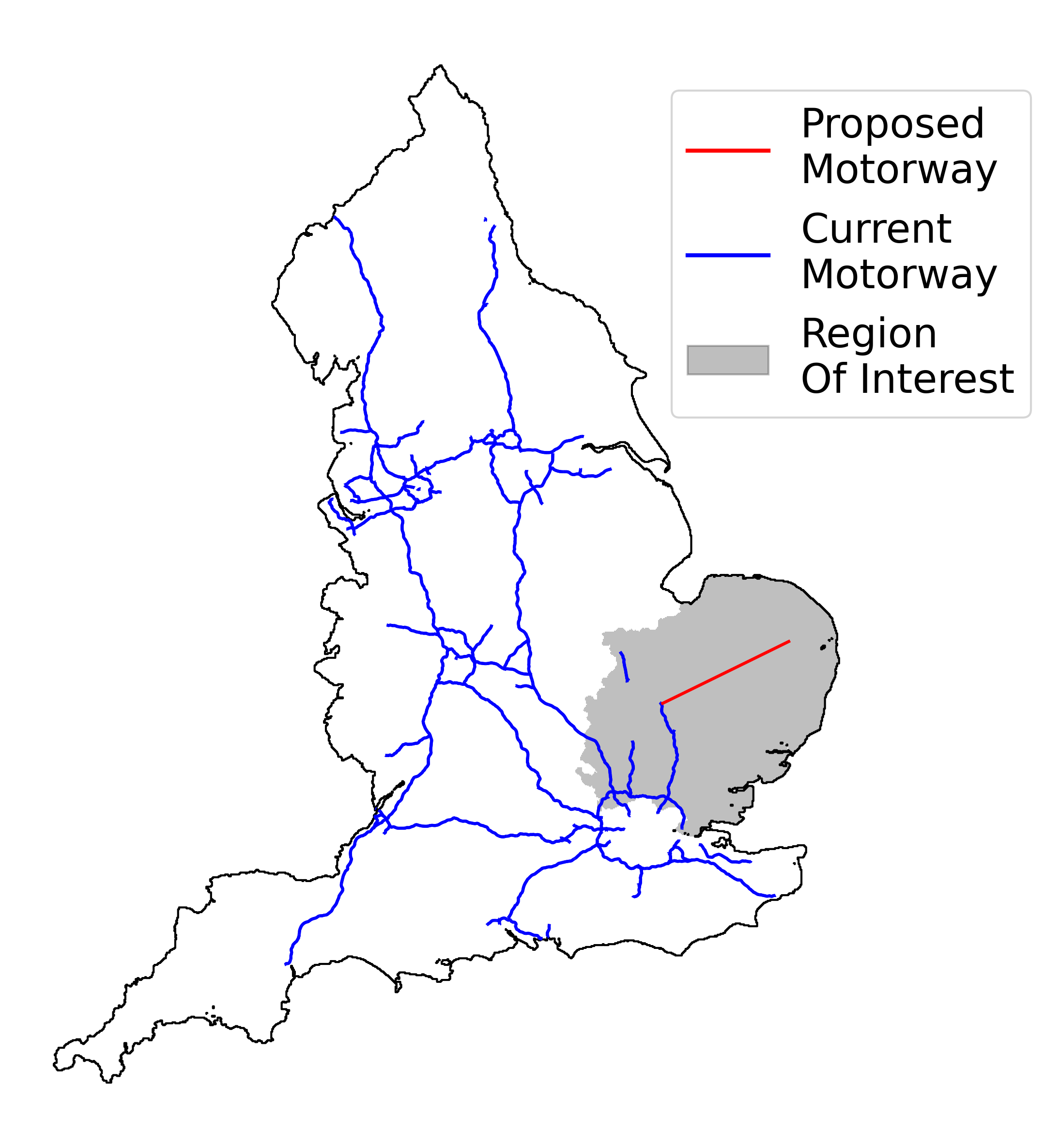}
  \end{subfigure}%
  }
  \hspace*{\fill}   % maximize separation between the subfigures
  \raisebox{10mm}{
  \begin{subfigure}{0.25\textwidth}
    \includegraphics[width=\linewidth]{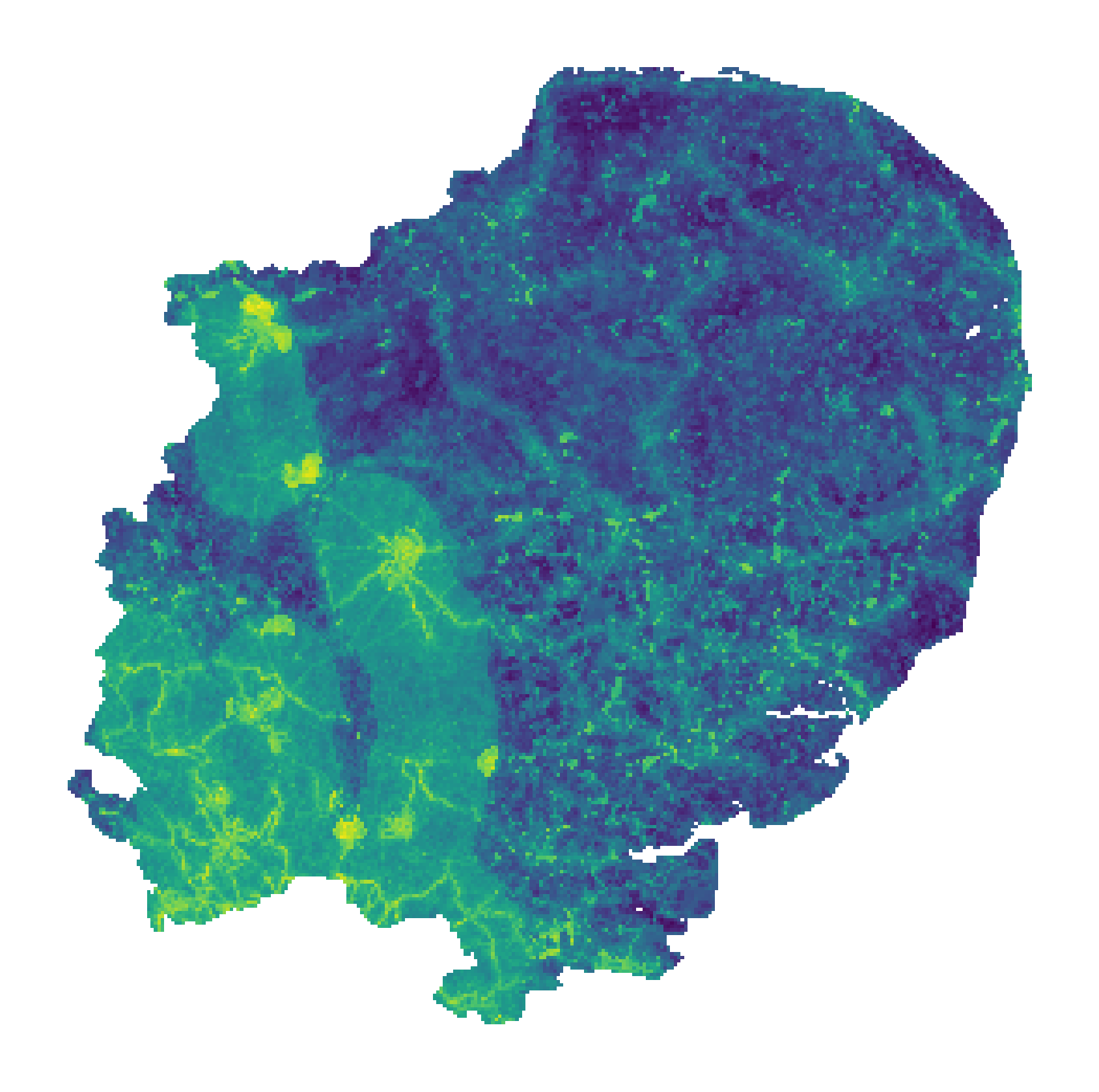}
    \caption{Current NO$_2$} \label{fig:policyHardCurrent}
  \end{subfigure}%
  }
  \hspace*{\fill}   % maximize separation between the subfigures
  \raisebox{10mm}{
  \begin{subfigure}{0.25\textwidth}
    \includegraphics[width=\linewidth]{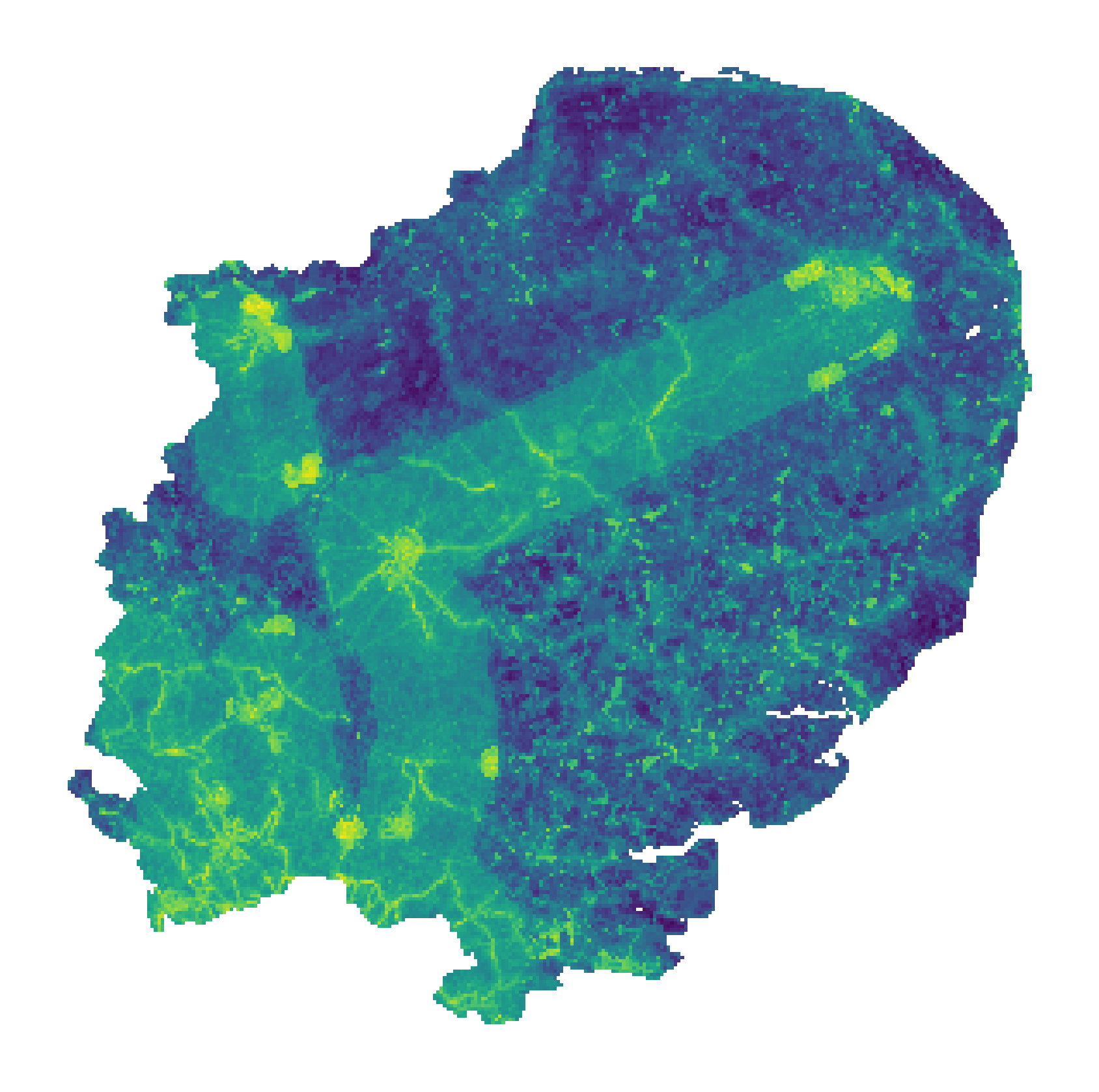}
    \caption{Predicted Future NO$_2$} \label{fig:policyHardFuture}
  \end{subfigure}%
  }
  \hspace*{\fill}   % maximize separation between the subfigures
  \raisebox{25mm}{
  \begin{subfigure}{0.08\textwidth}
    \includegraphics[width=\linewidth]{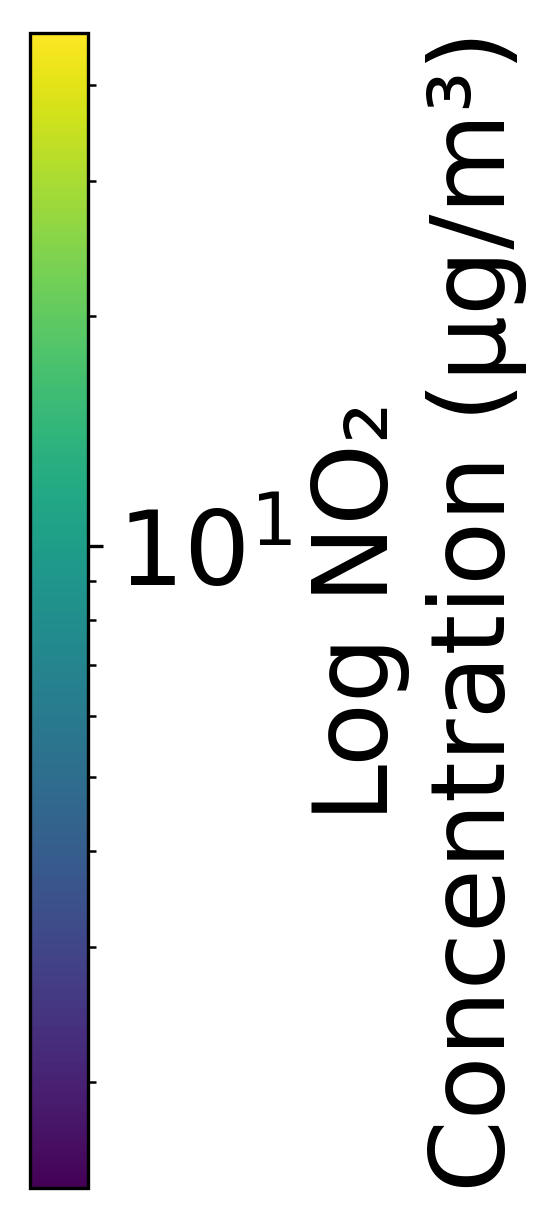}
  \end{subfigure}%
  }
\caption{{\bfseries Changes in NO$_2$ air pollution concentrations due to the placement of an additional motorway segment.} The spatial map of the NO$_2$ concentrations highlights the impact of placing an additional motorway segment on the local area. Of note is that the placement of the motorway segment will impact other air pollutants, shown in Figure \ref{fig:hardInterventionLineConcentrations}. } \label{fig:hardInterventionMotorwaySegment}
\end{figure}

\begin{figure}
    \hspace*{\fill}   % maximize separation between the subfigures
  \begin{subfigure}{0.49\textwidth}
    \includegraphics[width=\linewidth]{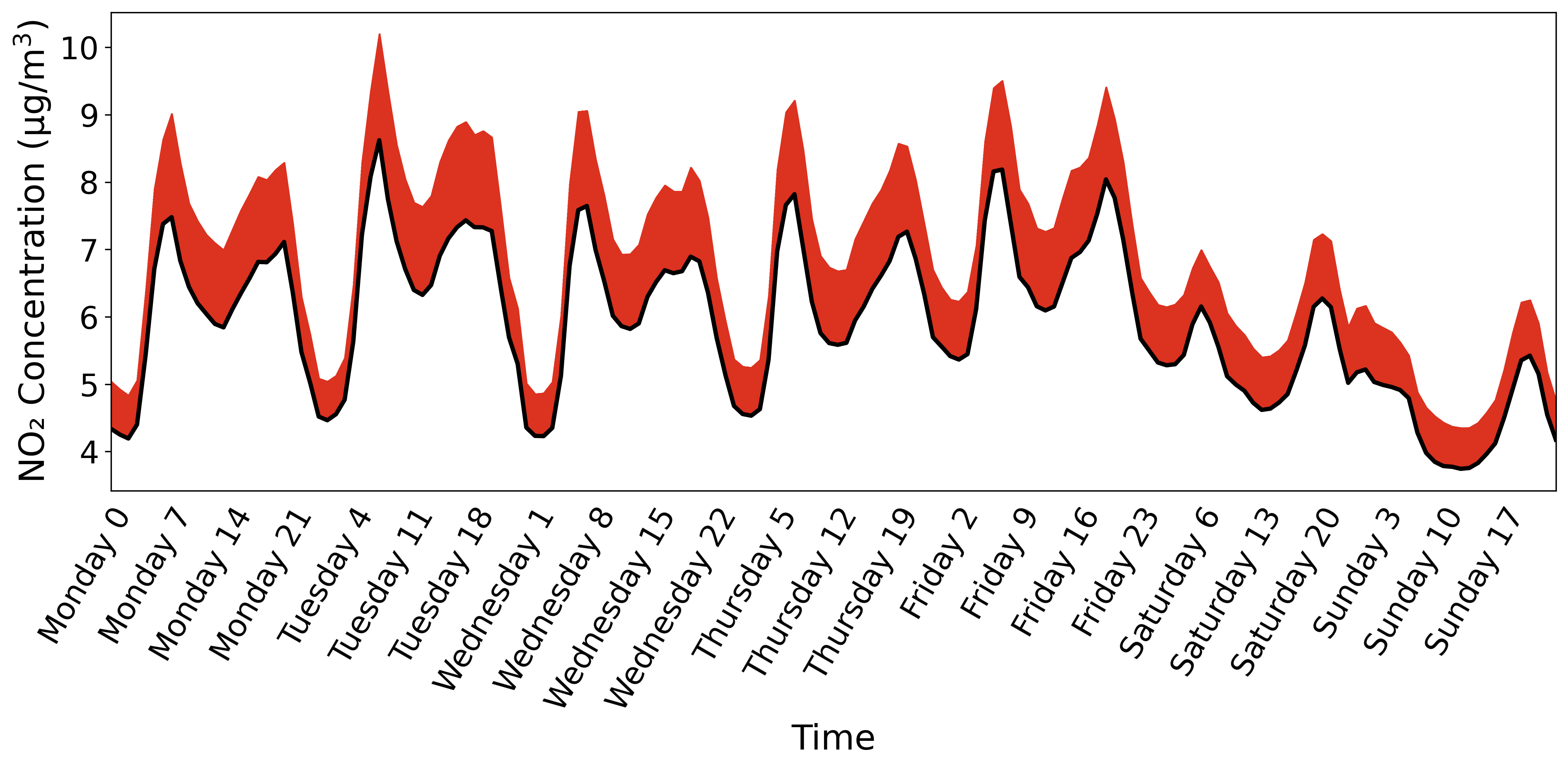}
    \caption{NO$_2$} \label{fig:policyHardLineNO2}
  \end{subfigure}%
  \hspace*{\fill}   % maximize separation between the subfigures
  \begin{subfigure}{0.49\textwidth}
    \includegraphics[width=\linewidth]{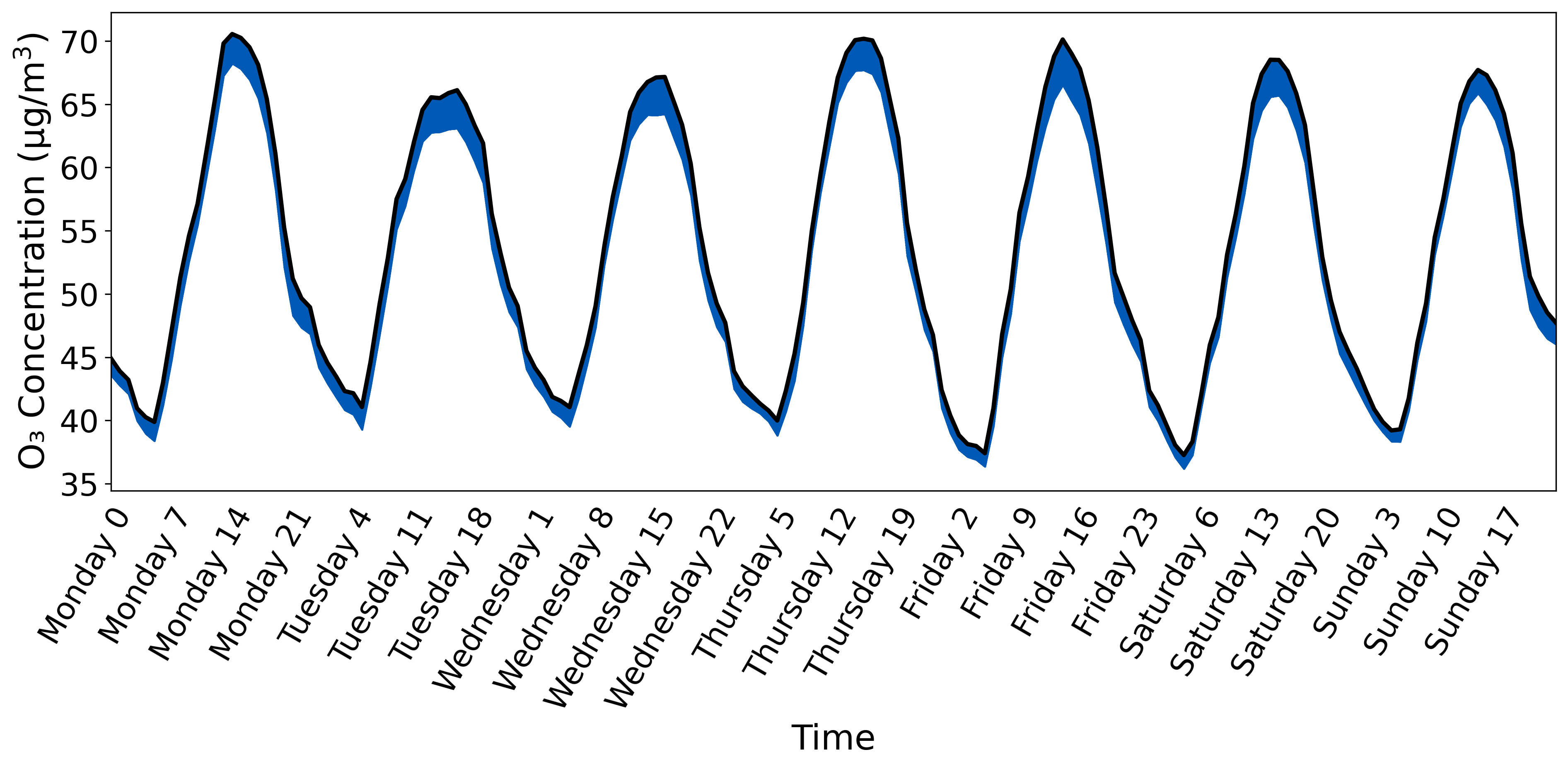}
    \caption{O$_3$} \label{fig:policyHardLineO3}
  \end{subfigure}
  \hspace*{\fill}   % maximize separation between the subfigures
  \\
  \hspace*{\fill}   % maximize separation between the subfigures
  \begin{subfigure}{0.49\textwidth}
    \includegraphics[width=\linewidth]{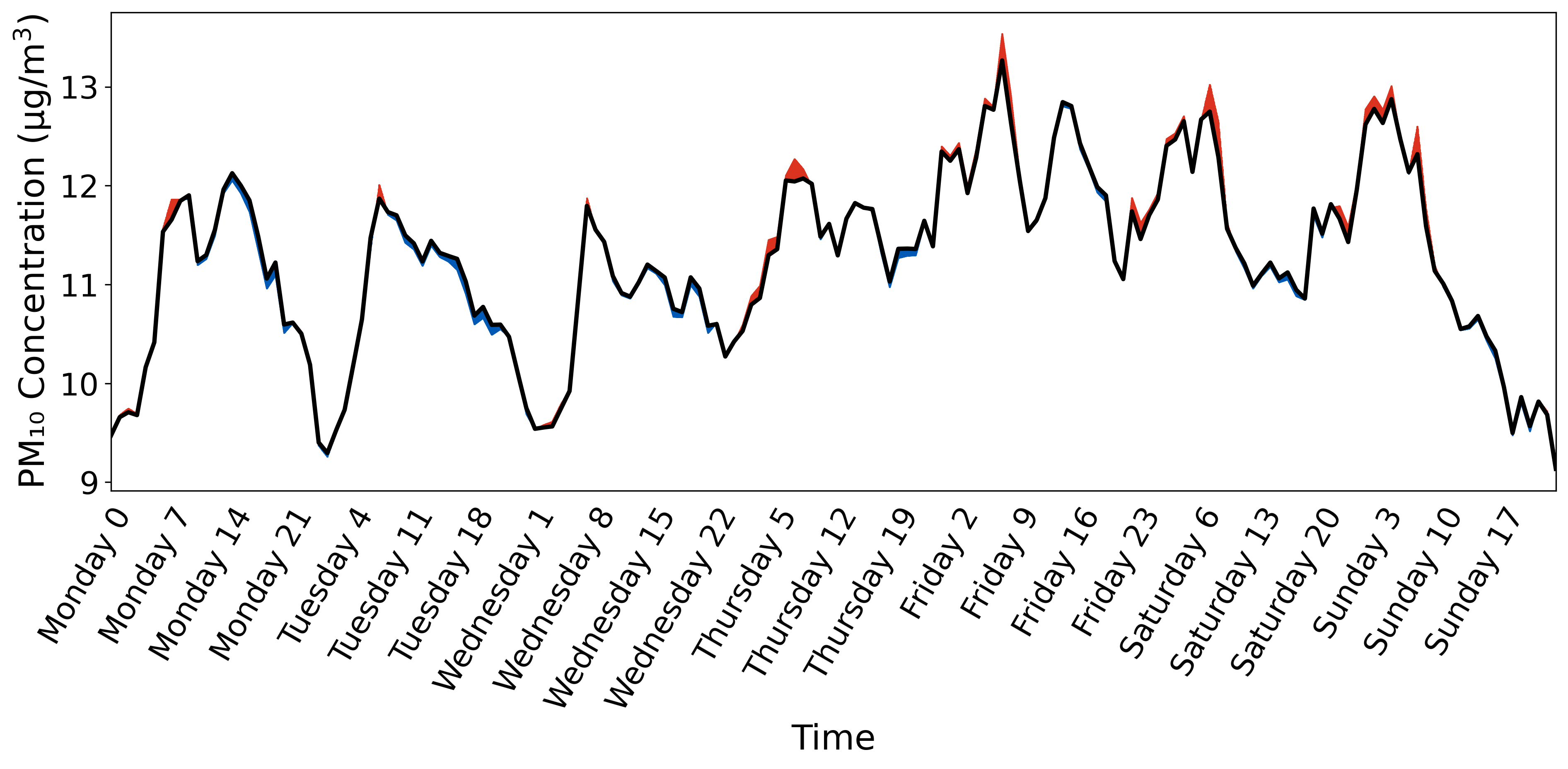}
    \caption{PM$_{10}$} \label{fig:policyHardLinePM10}
  \end{subfigure}%
  \hspace*{\fill}   % maximize separation between the subfigures
  \begin{subfigure}{0.49\textwidth}
    \includegraphics[width=\linewidth]{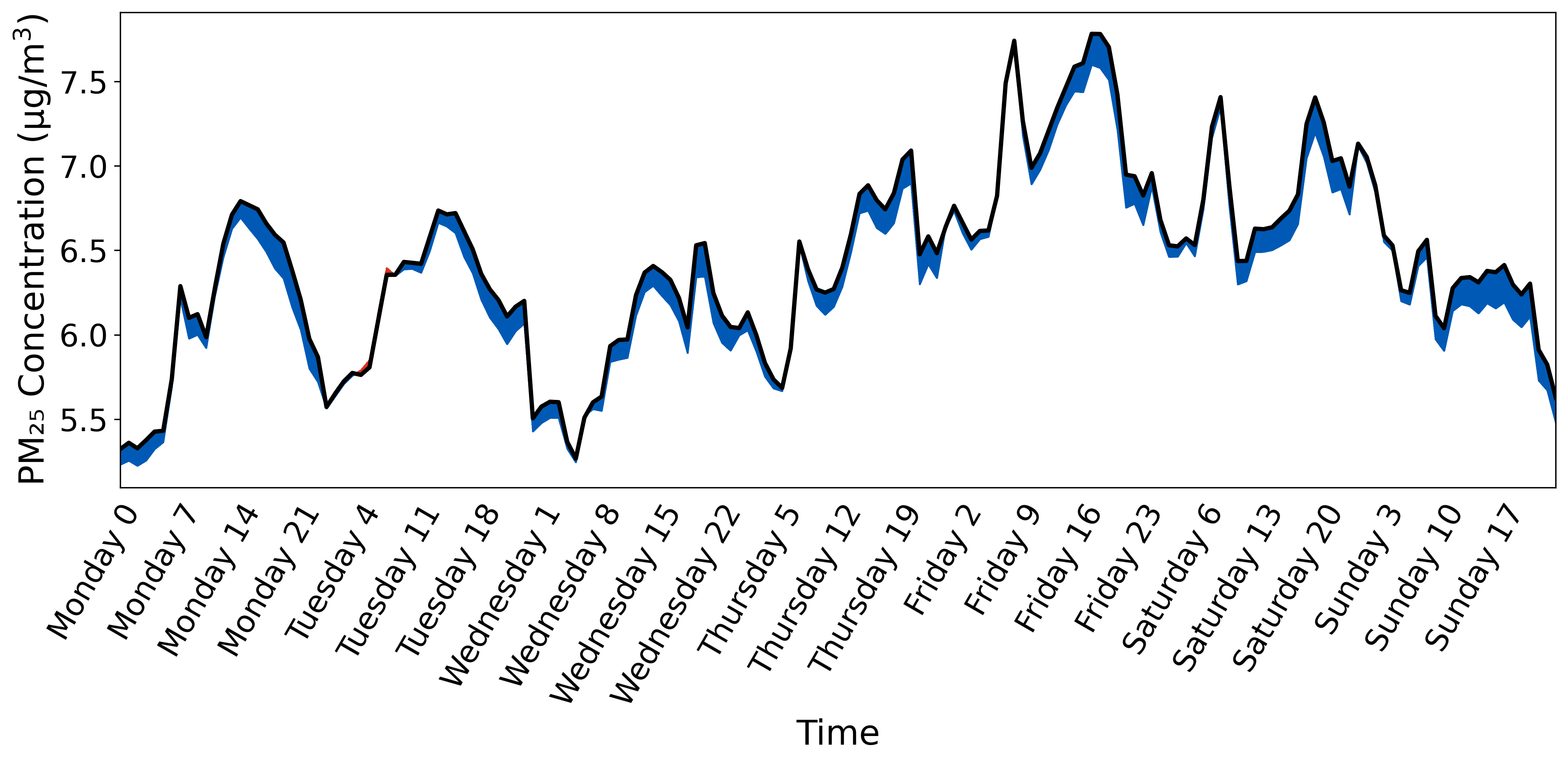}
    \caption{PM$_{2.5}$} \label{fig:policyHardLinePM25}
  \end{subfigure}
  \hspace*{\fill}   % maximize separation between the subfigures
  \\
  \hspace*{\fill}   % maximize separation between the subfigures
  \begin{subfigure}{0.49\textwidth}
    \includegraphics[width=\linewidth]{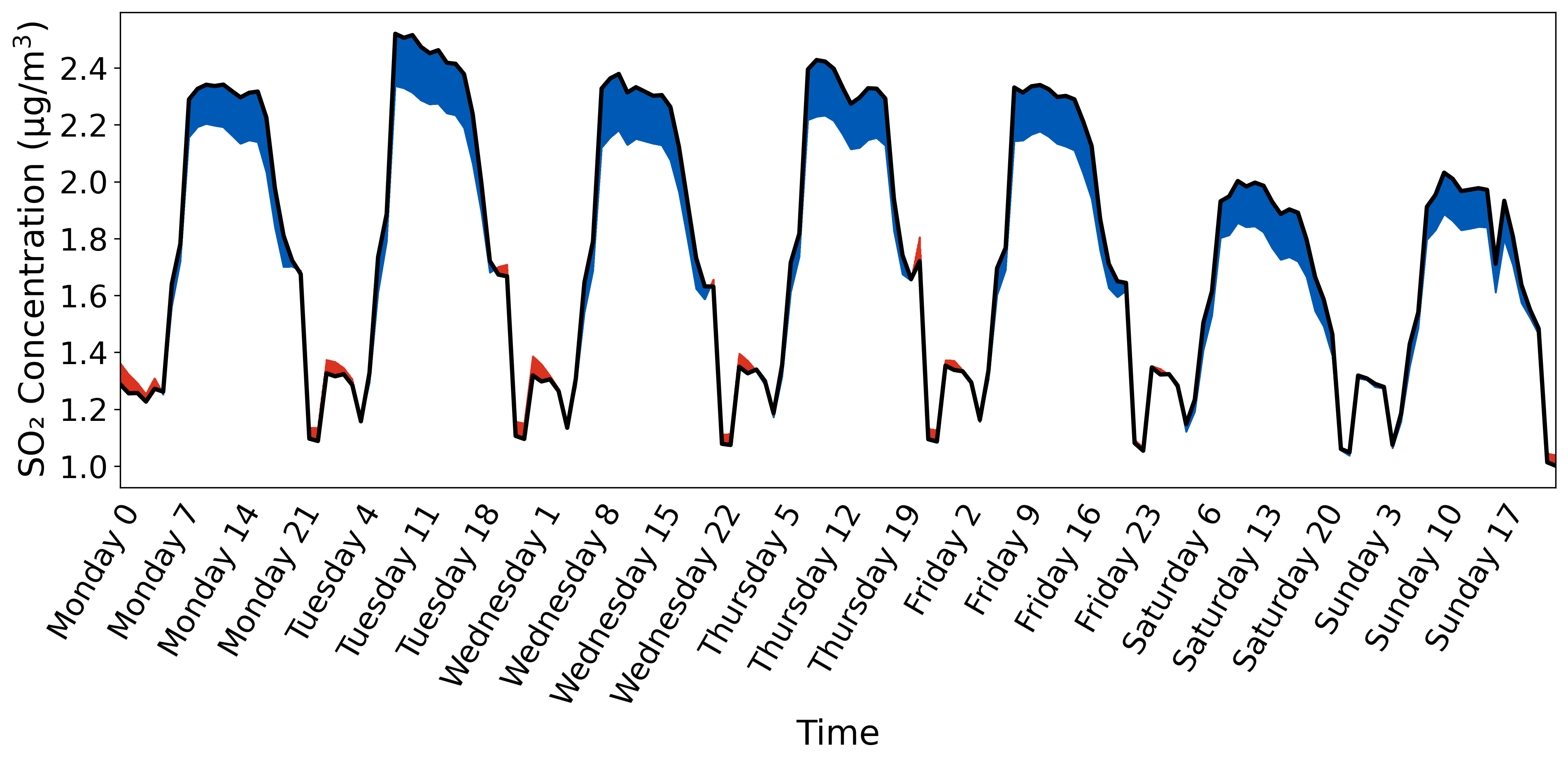}
    \caption{SO$_2$} \label{fig:policyHardLineSO2}
  \end{subfigure}
  \hspace*{\fill}   % maximize separation between the subfigures
  \raisebox{24.5mm}{
  \begin{subfigure}{0.25\textwidth}
    \includegraphics[width=\linewidth]{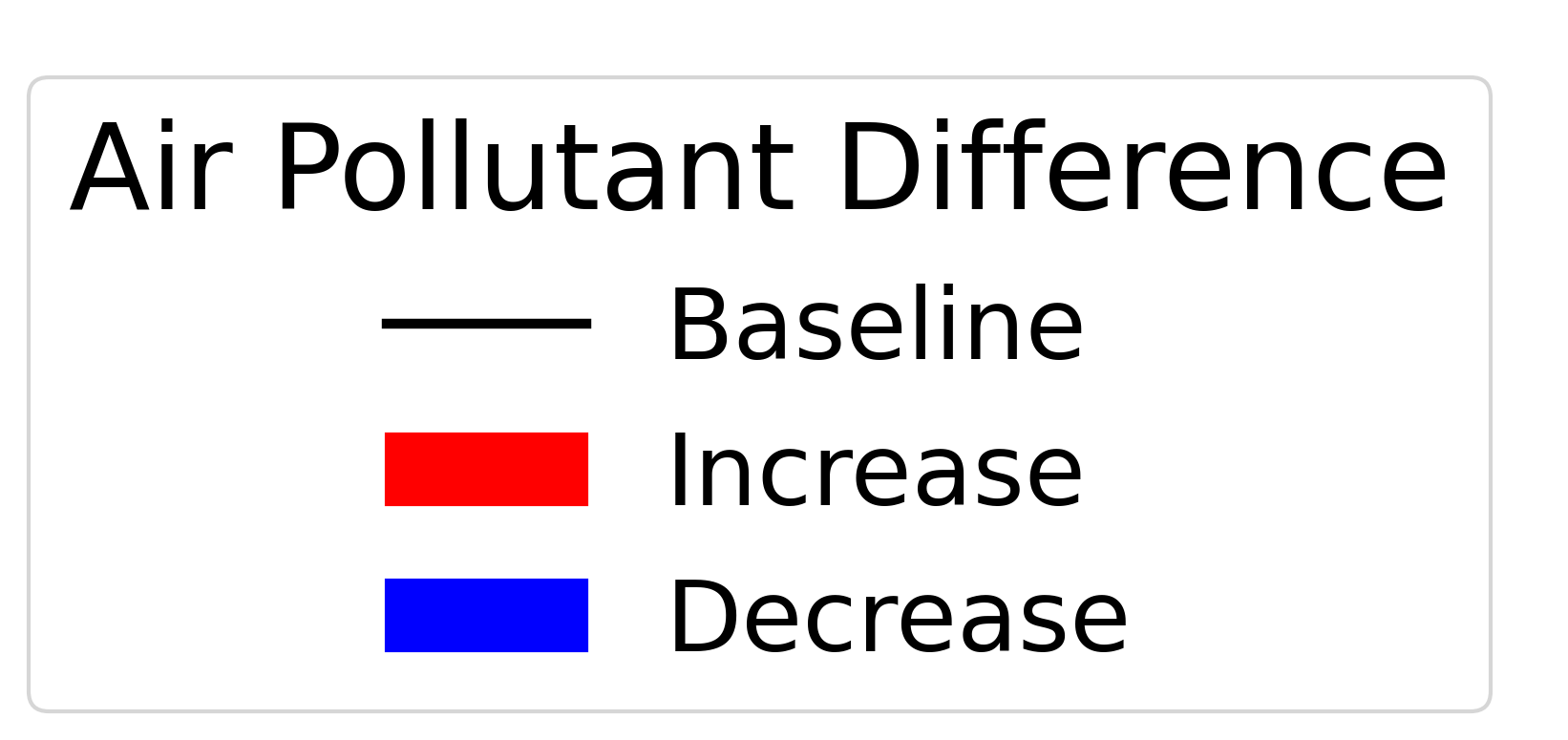}
  \end{subfigure}
  \hspace*{\fill}   % maximize separation between the subfigures
  }
  \hspace*{\fill}
\caption{{\bfseries Air pollution concentrations mean change after placing the additional motorway segment shown in Figure \ref{fig:hardInterventionMotorwaySegment}.} While NO$_2$ increases at all timesteps, the reverse is true for O$_3$, which sees a concentration reduction. In contrast, other air pollutants have more complex relationships, such as PM$_{10}$ increasing during the day but decreasing at night. } \label{fig:hardInterventionLineConcentrations}
\end{figure}

An example of a hard intervention the proposed package can support is the placement of new infrastructure, such as a motorway segment. Figure \ref{fig:hardInterventionMotorwaySegment} illustrates using the package to understand the impact of constructing a new motorway segment in the East of England region. Highlighted are the changes to the region's NO$_2$ concentrations. While an increase in NO$_2$ concentration due to the additional motorway segment is anticipated, the package quantifies these increases. Furthermore, the package offers insights into changes in different air pollutants, as depicted in Figure \ref{fig:hardInterventionLineConcentrations}, with some air pollutants such as O$_3$ decreasing, and others like PM$_{10}$ exhibiting a complex relationship with increases at some time points and decreases at others.

\noindent\textbf{Hard Intervention Functions}
\begin{itemize}
    \item Function \textit{get\_highways\_as\_geodataframe}  Fetch highways of a specified type within a bounding box from OpenStreetMaps and return as a GeoDataFrame.
    \item Function \textit{ckd\_nearest\_LineString} Calculate the nearest points between two GeoDataFrames containing LineString geometries.
    \item Function \textit{get\_even\_spaced\_points} Generate a list of evenly spaced points between two given points.
    \item Function \textit{calculate\_new\_metrics\_distance\_total} Simulate the addition of a proposed highway to current infrastructure and calculate new metrics.
\end{itemize}

\section{Discussion}

The work presented fills a gap in the current offerings of air pollution software by providing a single, easy-to-use, computationally fast Python package for accessing, visualising, and predicting air pollution concentration data. Furthermore, the available data is at a spatial and temporal resolution not previously publicly accessible, at an hourly 1km$^2$ resolution in England, and hourly 0.25$^{\circ}$ resolution globally. Alongside the data, tools for visualising key components of air pollution datasets are presented, empowering stakeholders to create high-quality visualisations of specific air pollution datasets to help engage others and promote their agenda. Finally, the package provides access to a lightweight, data-driven, supervised machine learning model to enable stakeholders to explore air pollution futures with environmental conditions important to them, whether that be exploring a new motorway placement, changing meteorological conditions due to climate change, or increased emissions from factories in a specific sector of the economy. The Environmental Insights package ensures that a problem as significant as air pollution is truly democratised for those who need it.

This work has introduced a method for enabling responsive two-way engagement with air pollution research conducted by the authors. The package offers functions for conducting analysis and an example workflow for creating the figures presented in this work. Additionally, the GitHub page for the package serves as an open forum for two-way communication between air pollution model creators, such as ourselves, and the broader community that consumes and utilises the outputs. This facilitates steering the research in directions necessary to support stakeholders in the manner they require.

\clearpage
\bibliographystyle{IEEEtran}
\bibliography{refs}

\end{document}